\documentclass[twocolumn,10pt, reprint, nofootinbib, amsmath, amssymb, aps]{revtex4-2}

\usepackage{graphicx}
\usepackage{dcolumn}
\usepackage{bm}
\usepackage{amssymb}
\usepackage{hyperref}
\usepackage{multirow}
\usepackage{color}

\begin{document}

\title{Phase Separation Kinetics and Cluster Dynamics in Two-Dimensional Active Dumbbell Systems}

\author{C. B. Caporusso$^{1}$, 
L.~F. Cugliandolo$^{2,3}$
P. Digregorio$^{4,5}$,
G. Gonnella$^1$, 
A. Suma$^{1}$,
}
\affiliation{
 $^1$Dipartimento Interateneo di Fisica, Universit\'a degli Studi di Bari and INFN, Sezione di Bari, via Amendola 173, Bari, I-70126, Italy, \\
 $^2$Sorbonne Universit\'e, Laboratoire de Physique Th\'eorique et Hautes Energies, CNRS UMR 7589, 4 Place Jussieu, 75252 Paris Cedex 05, France\\
 $^3$Institut Universitaire de France, 1 rue Descartes, 75231 Paris Cedex 05, France, \\
 $^4$Departement de Fisica de la Materia Condensada, Facultat de Fisica, Universitat de Barcelona, Mart\'{\i}  i  Franqu\`es  1,  E08028  Barcelona,  Spain, \\
 $^5$UBICS  University  of  Barcelona  Institute  of  Complex  Systems,  Mart\'{\i}  i  Franqu\`es  1,  E08028  Barcelona,  Spain
}

\begin{abstract} 
  Molecular dynamics simulations were employed to investigate the phase separation process of a  
  two-dimensional active Brownian {dumbbell} model. We evaluated the time dependence of the 
  typical size of the dense component using the scaling properties of
  the structure factor, along with the averaged number of clusters and their radii of gyration. The
  growth observed is faster than in active  particle (disk) models, and this effect is further enhanced under stronger activity. 
  Next, we focused on studying the hexatic order of the clusters.  The length associated to the orientational order grows algebraically and faster than for active Brownian particles. Under weak active forces, most clusters exhibit a uniform internal orientational order. However, under strong forcing, large clusters consist of domains with different orientational orders. We demonstrated that the latter configurations are not stable, and given sufficient time to evolve, they eventually reach homogeneous configurations as well. No gas bubbles are formed within the clusters, 
 even when there are patches of different hexatic order.
   Finally, attention was directed towards the geometry and motion 
 of the clusters themselves. By employing a tracking algorithm, we showed 
  that clusters smaller than the typical size at the observation time exhibit regular shapes, 
  while larger ones display fractal characteristics. In between collisions or break-ups, the clusters behave 
  as solid bodies. Their centers of mass undergo circular motion, with radii increasing with the cluster size. 
  The center of mass angular velocity equals that of the constituents with respect to their center of mass. 
  These observations were rationalised with a simple mechanical model.
\end{abstract}

\maketitle
\noindent
\tableofcontents

\section{Introduction}

Active matter refers to a class of systems composed of entities that can convert stored or ambient energy into directed motion~\cite{bechinger2016,Ramaswamy10,Vicsek12,Marchetti13,gonn15,Elgeti15,Care2019,Gompper_2020}. 
These particles or agents can exhibit self-propulsion or autonomous motion. 
Active matter systems can exhibit complex dynamic behaviors~\cite{sanchez2012,demagistris2015,cagnetta2017large,kumar2018,Doostmo2018,Santillan2018,negro2019hydrodynamics,carenza2020soft,giordano2021activity,Gompper_2020,Fodor23,Dauchot23,Caporusso-chiral23,Wiese23,Caprini23,Bayram23} due to the internal energy conversion processes of their constituents.
Examples are varied and cover natural and artificial systems at very different scales, 
including motor proteins in cells~\cite{butt} and flocks of birds~\cite{ballerini} on the one hand and active colloids~\cite{Palacci10,Lowen} or granular matter~\cite{Narayan07,kudrolli2008swarming} on the other.

The physicists' goals in this field are  to understand the principles governing their collective behaviors
and eventually explore potential applications in robotics, materials science or other.
In particular, much emphasis is set on establishing their phase behavior and
dynamic properties, and whether these are generic, in the sense of their global features 
not depending much on details such as the form of the elements and/or the type of 
microscopic dynamics. 

In this work we will study, with numerical methods,  the approach to stationarity of an Active Brownian Dumbbell (ABD) system 
with parameters such that an homogenous initial state tends to phase separate and form clusters~\cite{Suma14,Cugliandolo2017,Petrelli18}. We
will compare the dynamic behaviour of this system to the one of Active Brownian Particles  (ABP)
which has been already analysed in great detail in the literature~\cite{Digregorio19,Fily12,Redner13,negro2022hydrodynamic,Caporusso20,Caporusso22}. Qualitative and quantitative
similarities and differences will be stressed.

The article is organised as follows. In Sec.~\ref{sec:background} we will present background information.
Section~\ref{sec:model} introduces the model and numerical methods.
In Sec.~\ref{sec:growth-component} we analyse the growth of the dense component
and in Sec.~\ref{sec:orientational} the one of its internal orientational order.
Section~\ref{sec:clusters} investigates the motion of the clusters and develops a simple 
mechanical model that rationalizes it.
Finally, in Sec.~\ref{sec:conclusions} we present a summary of our results
and conclusions.

\section{Background}
\label{sec:background}

\vspace{0.2cm}

\noindent
        {\it The influence of the shape of the constituents.}

\vspace{0.2cm}

        The behavior of anisotropic active particles is generically richer than that of their isotropic counterparts~\cite{Moran22}. Some specific features of 
        these systems which
        have been discussed in the recent literature are the following.
        (i) The pressure calculated from mechanical, virial, and thermodynamic routes do not necessarily coincide.
        In the mechanical version,  torque interaction between the constituents and the confinement may render this {\it active pressure}
        a boundary-dependent property~\cite{solon2015pressure,Fily18,Pirhadi21}.  Moreover,
        the pressure exerted by an under-damped dumbbell system depends on the damping coefficient~\cite{Joyeux16,Joyeux17}.
        (ii)  Local nematic and even polar ordering can set in. The axial form of the dumbbells 
	allows for the alignment of head-tail dumbbell orientations~\cite{Suma14,Petrelli18}.
        (iii)
        At sufficiently strong activity dumbbells aggregate and the clusters
        exhibit coherent and long-lived rotations due to steric interactions which essentially quench the dumbbell's
        polarisation within the dense droplets. A distinguishing aspect of
        the rotating aggregates is that the velocity and
        orientational patterns are distinct: the former is a vortex, the latter a spiral~\cite{Caporusso23}.
        
On the experimental side, colloidal particles with non-convex shapes 
have been proposed to serve as the  building
blocks for active colloidal molecules with a dynamical
function~\cite{Lowen18,Mallory18,Paddy23}.

\vspace{0.2cm}

\noindent
        {\it Phase diagrams.}

\vspace{0.2cm}

        The collective behavior and phase diagrams of macroscopic systems with different constituents' shapes
        can be  rather different.
        Active Brownian Particles (ABP) placed in two dimensional spaces and interacting repulsively 
undergo Motility Induced Phase Separation (MIPS)~\cite{Cates15,Gonnella15,Digregorio18}
        when the activity is larger than a critical value. Two-dimensional rigid and repulsive 
dumbbell systems, instead,
        phase separate all the way down to zero activity~\cite{Cugliandolo2017} when the active force acts along their main axis.
        Shape and force anisotropy can combine to produce critical densities both lower and higher than those of disks~\cite{Moran22}.
        Differently from disk and dumbbell, 
rods can slide past each other and do not undergo MIPS~\cite{Peruani2006,Ginelli2010,Yang2010,Dijkstra19,Bar20}.

\vspace{0.2cm}

  \noindent
  {\it The active dumbbell liquid.}

\vspace{0.2cm}

        The stationary dynamics of repulsive dumbbells at not too high densities, so that  
the samples remain homogeneous, were studied in quite some detail~\cite{Suma14b,Suma14c}.
In these papers, the mean-square displacement, linear response function, and deviation from the equilibrium fluctuation-dissipation theorem were characterized as a function of activity strength, packing fraction, and temperature.
        The dynamics of passive thermal (in contact with the thermal bath) 
and athermal (disconnected from it) tracers interacting repulsively with 
the elements of active dumbbells' dilute baths were analyzed in~\cite{Suma16}. The point of this paper 
was to show for which parameters the dynamics of the tracers represent faithfully the ones of the dumbbells.

\vspace{0.2cm}

\noindent
{\it The phase separated regimes.}

\vspace{0.2cm}
       
        Rigid dumbbells cannot freely (at no energy cost) rotate within a crystallite without affecting its underlying structure.
        More precisely, in the clusters and even at their surface, the dumbbells get stuck and this  facilitates phase separation.
        Thus, aggregates made of dumbbells behave quite differently from those made of disks, when particle self-propulsion
        is switched on. In consequence, clusters not only translate but also rotate with, on average,
        an angular velocity that decays proportionally to their inverse size~\cite{Schwarz-Linek12,Tung16,Petrelli18}
        but also depends on their shape.


In the phase separated region of the phase diagram, 
the dilute phase has no order and behaves as an active
liquid or gas. Instead, the dense phase has local orientational order, 
in the sense that the modulus of the local hexatic order parameter takes a large
value~\cite{Cugliandolo2017,Petrelli18}.
The axial form of the molecules allows for the development of polar order, which turns out to change 
with Pe:
no polarization is found at small Pe, the dumbbells within the clusters orient along the radial 
direction for intermediate Pe, and the clusters show a spiralling structure at large Pe~\cite{Petrelli18}. 

Similarly to what was measured for Active Brownian Particle systems within MIPS~\cite{Petrelli20}, 
the motion and linear response of the dumbbells  confined to clusters or freely displacing in the dilute phase
are expected to be very different.

\vspace{0.2cm}

\noindent
{\it The cluster's motion.}

\vspace{0.2cm}

At sufficiently large Pe the clusters not only translate but can also rotate~\cite{Suma14,Petrelli18}. 
Each particle exerts a local torque which is balanced by the drag on the cluster, sustaining rotations.
For Pe $\geq$ 50, the spontaneously formed rotating clusters turn around their center of mass 
with an angular velocity that is proportional to the inverse of their radii. 

A careful analysis of this kind of motion
was carried out in~\cite{Petrelli18} were, in particular, the dependence of the 
enstrophy on  Pe was compared to the one of the kinetic energy. While the latter measures
the strength of flow in the system, the former measures the presence of vortices in the velocity field.
For high values of the activity  the probability distribution function of the 
enstrophy has a multi-peak structure. These peaks progressively disappear for decreasing values of Pe. 
Below Pe $\sim$ 40 
the probability of finding non-vanishing enstrophy is almost zero.

\vspace{0.2cm}

\noindent
{\it Other aspects.}

\vspace{0.2cm}

  \noindent
        Other aspects of two dimensional dumbbell models were addressed in the literature.         
        Dilute suspensions of active Brownian dumbbells interacting with competing short attractive and long range repulsive potentials,
        a potential that mimics the interaction of weakly charged particles in the presence of depletants,
        for different degrees of particle activity were studied in~\cite{Tung16}.
        The dynamics of flexible active Brownian dumbbells with and without shear flow was addressed in~\cite{Winkler2016a, Coples}.
        Other studies include
        tests of Green-Kubo relations in a chiral active dumbbell fluid~\cite{Hargus20} and the analysis of  similarities and differences between active and passive
        glasses made of dumbbell molecules~\cite{Mandal17,Mandal2018}.
        The influence of correlations in the propulsion direction was studied in~\cite{Siebert2017}, where
        systems of dumbbells joined by an elastic string, with forcing in uncorrelated direction,  or correlated direction but not necessarily the axial one, were simulated.
        In the former case phase separation is pushed to higher values of Pe since part of the propulsion energy is spent in stretching the elastic bond between the disks.

We close this introduction by mentioning that 
the study of three dimensional active dumbbell system is only starting.
The dynamic phase diagram of a (rather dilute) active attractive dumbbell model was recently established in~\cite{Caporusso23}, where the motion of
a dumbbell cluster was also characterized. The phase separation in a mixture of “hot” and “cold” three-dimensional dumbbells was analyzed in~\cite{Venkatareddy23}.

 \vspace{0.2cm}
 
  \noindent
  {\it The kinetics of phase separation}
  
  \vspace{0.2cm}
  
        Although quite a lot is known about the dynamics of active dumbbell systems in their steady states,
        the kinetics of phase separation has been much less considered~\cite{Suma13,Suma14,Digregorio19}.
        In~\cite{Suma13} the early stages of growth were analyzed (second regime  identified below, before entering the proper scaling limit).
        A growing length $R(t)\sim t^a$ was measured from  the inverse of the first moment of the
        spherically averaged structure factor, with $a=0.90$ at $\phi=0.4$ and $a=0.65$ at $\phi=0.6$,
        both at Pe = 200.

In this work we investigate the dynamics of formation of the dense phase, 
carrying out an analysis that parallels the ones performed in~\cite{Caporusso20,Caporusso22}  for ABPs. 

Compared to previous studies of growth in {Active Brownian Dumbbell (ABD)}  systems, we use 
much larger system sizes, which let us to properly control the finite size effects, 
and we run the simulations over much longer time scales, which allow us to explore 
in detail the  scaling regime.

\section{Model and numerical methods}
\label{sec:model}

In this Section we describe the model and the numerical techniques that we used to
study it. We also present a very short summary of its phase behavior, around
the phase diagram shown in Fig.~\ref{fig:phase-diagram}.

\subsection{Model}

We consider a two-dimensional system of $N$ dumbbells.
Each dumbbell is a diatomic molecule,
with two identical disks rigidly connected together, for a total of $2N$ disks.
Each disk has diameter $\sigma_{\rm d}$ and
mass $m_{\rm d}$. The distance between
the two centers of the disks forming each dumbbell
is fixed and equal to $\sigma_{\rm d}$, in which case the dumbbells are also called
dimers.
We identify a tail and a head in each dumbbell,
which remain the same along the system's evolution,
and we apply the active force along this direction (see~\cite{Siebert2017} for variations
around these choices).

The disks, labelled by $i=1, \dots, 2N$
with center position ${\mathbf r}_i$, follow the Langevin equation of motion
\begin{equation}
  m_{\rm d} \ddot {\mathbf r}_i=-\gamma_{\rm d}\dot {\mathbf r}_i -{\boldsymbol \nabla}_i U+
  {\mathbf f}_{\rm act}+
  \sqrt{2k_BT\gamma_{\rm d}} \, {\boldsymbol \eta}_i \; ,
  \label{eq:bd}
\end{equation}
where
${\boldsymbol \nabla}_i=\partial_{\mathbf{r}_i}$, $\gamma_{\rm d}$ and $T$ are the friction coefficient and
the temperature of the thermal bath,
respectively,  and $k_B$ is the Boltzmann constant.
The last term in the right-hand-side is proportional to
${\boldsymbol \eta}_i$, a time-dependent
Gaussian white noise acting on each disk,
with vanishing mean, $\langle\eta_{ia}(t)\rangle=0$, and
independent delta correlations, $ \langle\eta_{ia}(t_1)\eta_{jb}(t_2)\rangle =\delta_{ij}\delta_{ab}\delta(t_1-t_2)$,
with $a,b=1,2$ the label for the two spatial coordinates. Henceforth we will indicate the
noise averages with angular brackets $\langle \dots \rangle$.
We note that the thermal noise affects the translational and rotational degrees of freedom
of the dumbbells.

We use the internal potential energy
\begin{equation}
  U= \sum_{i \neq j}^{2N}  U_{\rm Mie}(|{\mathbf r}_i-{\mathbf r}_j|)
  \; ,
\end{equation}
with the Mie potential~\cite{Mie1903}
\begin{equation}
  U_{\rm Mie}(r) = \left\{ 4\epsilon \left[ \left(\frac{\sigma}{r} \right)^{2n} - \left(\frac{\sigma}{r}\right)^{n} \right] +\epsilon \right\} \theta(2^{1/n}\sigma-r)
  \label{eqmie}
\end{equation}
and
$\theta$ the Heaviside function. The potential is truncated at its minimum ($r=2^{1/n}\sigma$) so that it is purely repulsive.
We set $n=32$
in order to have a very steep potential and as close as possible to the hard-disk limit without losing computational efficiency, 
$\sigma$ and $\epsilon$ set the length and energy scales of the potential.
We choose $2^{1/n}\sigma=\sigma_{\rm d}$ so that, on the one hand,  the minimum of the potential  equals
the disk diameter, and on the other hand, the derivative of the potential vanishes
at $r=\sigma_{\rm d}$. In this way, the {force} is continuous at this distance.

The potential does not account for the dumbbell's connectivity, which is taken care of with the RATTLE scheme (see numerical integration section),
which makes the bond completely rigid through an additional set of forces not described here.

The term ${\mathbf f}_{\rm act}$ in Eq.~(\ref{eq:bd})
represents the active force. It acts on the tail-to-head direction of each dumbbell and has constant modulus
$f_{\rm act}$. Differently from ABP models, 
dumbbells do not need an additional force in order to randomly rotate ${\mathbf f}_{\rm act}$
in the two-dimensional space. In fact, the thermal noises acting on the two disks forming each dumbbell
are independent, and their combination allows for the dumbbell's effective diffusive rotation.

We focus on the system's behavior dependence on two control parameters: the  surface fraction covered by the beads,
\begin{equation}
  \phi=\frac{N\pi\sigma_{\rm d}^2}{2L^2}
  \; ,
\end{equation}
with $L$ the linear size of the square box which contains the molecules,
and  the P\'eclet number,
\begin{equation}
  \mbox{Pe} = \frac{2\sigma_{\rm d} f_{\rm act}}{k_BT}
  \; ,
\end{equation}
which can be seen as
the ratio between the work done by the active force when translating the dumbbell by
its typical size, $2\sigma_{\rm d} f_{\rm act}$, and the thermal energy scale $k_BT$.
The larger Pe, the farther away from equilibrium the dynamics are. We keep all other parameters fixed.
We measure distances in units of the disks diameters.


\subsection{Numerical Integration}
\label{secnum}

We consider systems of different dumbbell numbers  $N=128^2/2$, $256^2/2$, $512^2/2$, $1024^2/2$ and $2048^2/2$.
The linear size $L$ of the box was set based on the target value of $\phi$, so that $L^2=N\pi\sigma_{\rm d}^2/(2\phi)$.

We initialize a system with  global density $\phi <0.6$ in a random initial condition,
assigning a random position to the
first bead's center of mass, and then a random angle to place the second one.
For higher densities, $\phi > 0.60$, we place the dumbbells in an ordered way and
let them disorder through a short run with no activity and high temperature.
We then let evolve these initial conditions with the Langevin dynamics in Eq.~(\ref{eq:bd}) at different Pe and $\phi$ values.

As customary, we use as system's units the mass $m_{\rm d}$, the diameter $\sigma_{\rm d}$ and the
typical potential energy $\epsilon$~\cite{allen}.
We then express all physical quantities using reduced units. For example, the
time unit is  $\sigma_{\rm d} (m_{\rm d}/\epsilon)^{1/2}$~\footnote{One could also use the rotational diffusion coefficient $D_r$ to define a different time unit as $\tau_r = 1/D_r$.}.
The thermal bath parameters are fixed to $\gamma_{\rm d}=10$ and $k_BT=0.05$ in  reduced units, with $k_B=1$.
The large $\gamma_{\rm d}$ value assures that the dynamics is close to over-damped.
Still, the numerical integration and measurement of all quantities presented in the paper are performed considering
the inertial term $m\ddot{\bold{r}}_i$. Typical simulations took between $10^5$ and $10^6$ simulation time units (MDs).

We used a velocity Verlet algorithm that solves  Newton's equations of motion, plus additional force terms for the Langevin-type  thermostat,
to numerically integrate the stochastic evolution equation.
We kept the  bonds  rigid with the help of the RATTLE scheme~\cite{rattle}.
This is equivalent to considering an additional force in Eq.~(\ref{eq:bd}), that  takes into account the holonomic constraints.
The time-step choice is related to the force
exerted during the simulation. We adapted it to enforce numerical stability.
In this paper, for systems at Pe $\leq 10$ we used a
time-step of $0.005$, while for Pe = 20 and Pe = 40 we used a time-step equal to $0.002$ and, finally,  for Pe = 100 and Pe = 200 the time-step was
reduced to $0.001$.

In order to efficiently parallelise the numerical computation
we used the open source software Large-scale Atomic/Molecular Massively Parallel Simulator \linebreak (LAMMPS), available
at {\tt github.com/lammps}~\cite{plimpton1995fast}.

The parameters Pe and $\phi$ considered lie
in the region of the phase diagram where phase separation occurs, i.e. where the system 
separate in dense and dilute components over time. In particular, we vary Pe and choose a packing fraction $\phi$ such that in the long time limit there will be a fixed proportion of dense and dilute phases, like 25\% - 75\%, 50\% - 50\%, or 75\% - 25\%. 

On average each simulation lasting $10^5$~MDs with $2N=256^2$ particles was run on 48 processors for a total of 50 hours for each CPU.
Typically, we collected data from  systems with $2N=256^2$ particles and we used $10-20$ independent runs to construct the averages and
probability distribution functions. In order to measure the growth exponent we ran simulations with $2N=2048^2$ particles on 
96 processors for $\sim 500 $ hours on each CPU, and we averaged over 5 independent runs.

\subsection{Cluster identification and tracking}
\label{subsec:cluster-identification}

We used a DBSCAN algorithm to identify clusters of dumbbells~\cite{Esler96}.
DBSCAN is a clustering algorithm, which distributes points into clusters according to the local point density.
We shortly outline hereafter the fundamental rules of the algorithm and we give the values of the parameters that we used. For more details about the algorithm
and its validation for the present problem, we refer to Ref.~\cite{Caporusso20}.
\begin{itemize}
  \item Given that two points are neighbors if their distance is less than a given extent $\varepsilon$, a point is a ``core point'' if it
        has at least $n_{\rm min}$ neighbors.
  \item Any two core points connected through a path in the neighbors' network belong to the same cluster, together with their neighbors.
  \item Points which are not cores and are not reachable from a core do not belong to any cluster.
\end{itemize}
We used $\varepsilon=1.5 \, \sigma_{\rm d}$ and $n_{\rm min}=6$ for a reliable identification of the clusters.

We also used the algorithm that we devised in Ref.~\cite{Caporusso22} to track the cluster's
trajectories individually. The method, carefully described in the SM of this reference,
is a combination of the DBSCAN routine, which identifies the clusters instantaneously at an
initial tracking time, and a routine to identify the same cluster at
two successive times. It should be noted, though, that
some clusters may be lost in the course of evolution. We will discuss this feature in the
analysis of the cluster dynamics.

\subsection{Phase diagram}

\begin{figure}[h!]
  \centering
  \includegraphics[width=\linewidth]{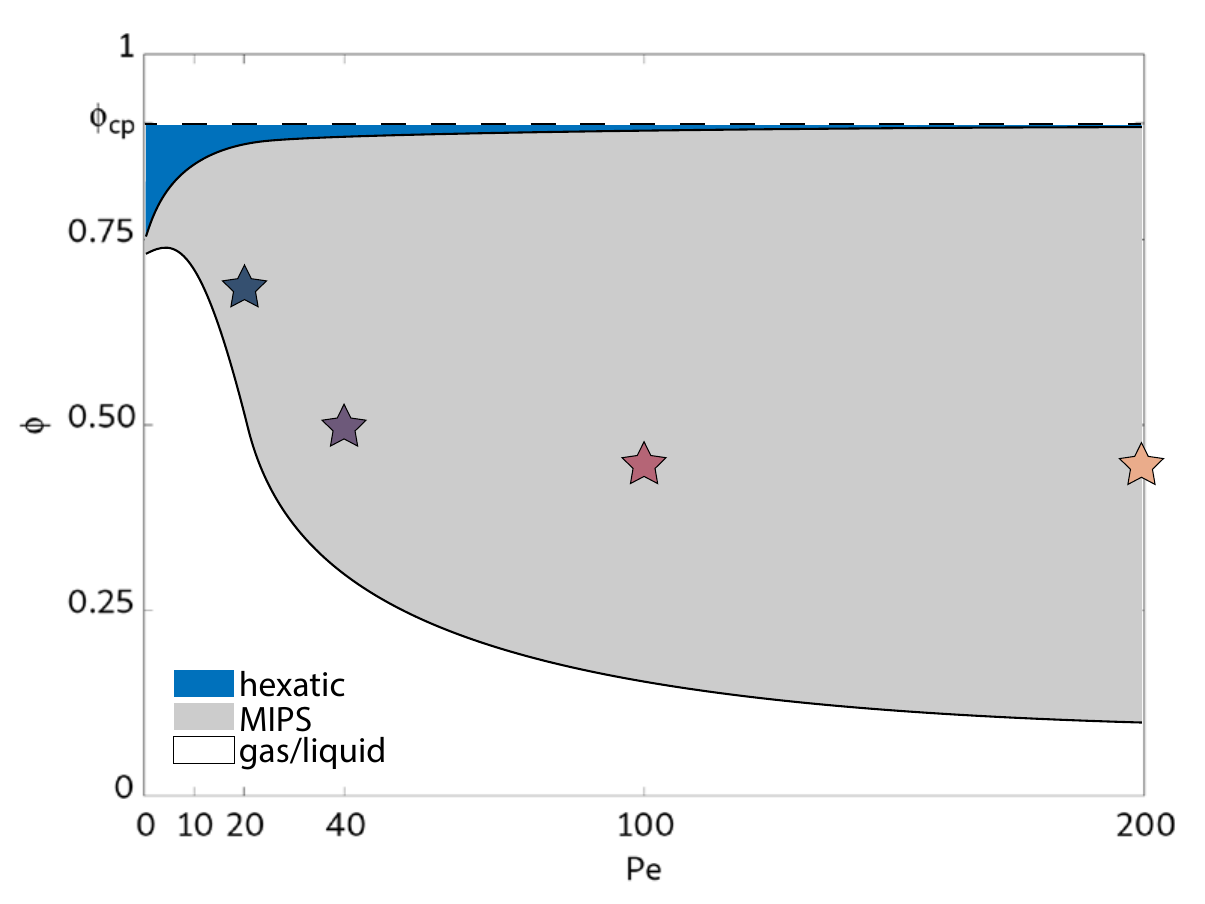}
  \caption{The  $\mbox{Pe}-\phi$ {\bf phase diagram} of the dumbbell system~\cite{Cugliandolo2017}.
    The stars locate parameters on the curve of a 50:50 mixture, i.e., in which the surface 
    occupied by the dense phase is half the total one. Most of the
    results that we present in the rest of the paper are for these parameters. 
    }
  \label{fig:phase-diagram}
\end{figure}

We study the dynamics of the dumbbell system in the phase separated region of the  phase diagram, established
in~\cite{Cugliandolo2017,Petrelli18,Digregorio19} under  similar conditions and
reproduced in Fig.~\ref{fig:phase-diagram}.
The  dimers beads that we consider have the same symmetry as disks do and, therefore,
their perfect crystalline order is also hexagonal, with closed packing fraction $\phi_{\rm cp} \sim 0.9$,
shown with a dashed horizontal line. The phases are represented with different colors and they correspond to the {gas/liquid} (white), hexatic (blue),
and phase separated regions (grey).  The colour code (online) will be the same in all figures. We note that
differently from the ABP system, in the dumbbell's one the co-existence region close to Pe = 0 extends continuously
to large Pe. There is no critical ending point for a Motility Induced Separated Phase (MIPS). The transition at
the lower critical line is discontinuous all along the curve. We did not identify a solid phase~\cite{Frenkel91,Frenkel93} above the hexatic one but
we cannot exclude that such a transition exists. We simply cannot  make reliable measurements at densities which are
near close packing.

\section{Growth of dense phase}
\label{sec:growth-component}

In this Section we discuss the way in which the dense component increases its size
starting from an initial random configuration. We also study the organization of the 
internal structure of this dense component. More precisely, we focus on the local hexatic (orientational) order.

\subsection{Growing length and dynamical regimes}

\begin{figure*}
  \centering
  \includegraphics[width=\textwidth]{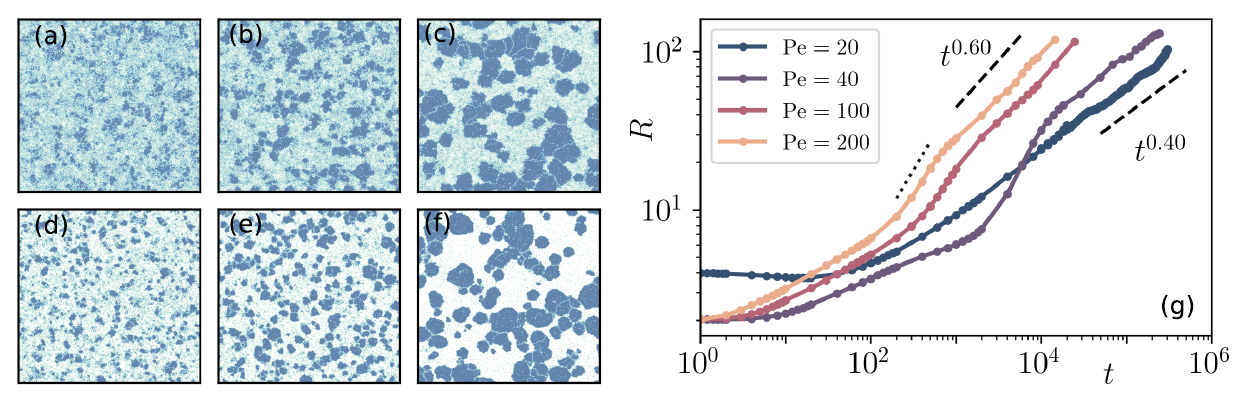}
  \caption{{\bf The phase separation process.}
  (a)-(f) Snapshots of the systems at Pe = 20 (first row) and Pe = 100 (second row), 
  at times such that the  average size of the dense component is the same (corresponding to an average density, respectively, $\phi = 0.70$ and $0.50$). 
  The instants at which the images were taken fall in 
  different regimes: $t = 2 \times 10^3$ for Pe = 20 (a) and $5 \times 10^2$ for Pe = 100 (d) are very short times, 
  $6 \times 10^3$ (b) and $10^3$ (e) fall in the rapid growth regime, finally $10^5$ (c) and $8 \times 10^3$ (f)  are in the scaling regime.
  {Light/dark blue color regions correspond to low/high local density regions}.
  (g) Grow length as a function of time for  parameters  corresponding to the stars in Fig.~\ref{fig:phase-diagram}: Pe = 20, 40, 100, 200 and the global packing fractions
  $\phi = 0.70, \ 0.50, \ 0.50, \ 0.450$, respectively, averaged over 5 independent runs.
  The short dotted segment represents the power 1 which is very close to the intermediate algebraic rapid growth. The dashed segments 
  show the power law fits  for the two extreme cases, Pe = 20 and Pe = 200, in the scaling regime.
  }
  \label{fig:Rt}
\end{figure*}

{In Fig.~\ref{fig:Rt}(a)-(f) we show snapshots of the evolution of a dumbbells system at Pe = 20 and 100 starting from a random conformation, in a phase-diagram region where the dense phase occupies 50\% of the surface at steady-state (long times). The time dependence of the typical length of the dense component $R(t)$  is}
calculated from the inverse first moment
\begin{equation}
  R(t) = \dfrac{\pi \int \, dk \; S(k,t)}{\int \, dk \; k \, S(k,t)}
  \label{eq:length}
\end{equation}
of the spherically averaged  disks centers' structure factor
\begin{equation}
  S({\mathbf k},t) = \frac{1}{2N} \sum^{2N}_{i} \sum^{2N}_{j} e^{i {\mathbf k} \cdot ({\mathbf r}_i(t)-{\mathbf r}_j(t))}
  \; .
\end{equation}
The radially symmetric $S(k,t)$ is the average over the reciprocal lattice vectors inside a spherical shell of thickness $2\pi/L$. {This growing length, $R(t)$,  for four Pe - $\phi$ pairs for  which there is a 50:50 
surface occupied by the dense and dilute phases
 in the steady state is plotted in Fig.~\ref{fig:Rt}(g). 
 We will later discuss in Fig.~\ref{fig:structure-factor} the structure factor for the same Pe - $\phi$ parameters.}  
We also considered the structure factor of the dumbbells' center of mass finding similar results
(not shown).

{We are able to distinguish three dynamic regimes in  $R(t)$ curves of Fig.~\ref{fig:Rt}(g) 
for Pe $\ge 40$, corresponding to different growth {stages} of the dense region.  For very short times (first regime) we see multi-nucleation  of  many small dense
droplets, see Fig.~\ref{fig:Rt}(d).  In the second regime there is a rapid growth of the typical length, corresponding to the growth of some droplets  by  condensation  and  aggregation, and evaporation of other droplets, see Fig.~\ref{fig:Rt}(e) and movies M1 and M2}~\footnote{Movies can be found at \url{www.dropbox.com/scl/fo/dhthjnhtpawy1xyr60k72/h?rlkey=3lpqd7e4zskm5hu2nb8ohoa8e&dl=0}}.
{In the third and final regime clusters grow using as main aggregation mechanism translation, collision and merging, eventually forming irregular and elongated structures (Fig.~\ref{fig:Rt}(f)), which will be discussed later in Sec.~\ref{subsec:geometry}. We will prove in Sec.~\ref{structfact} that this is a coarsening scaling regime. For now we will reference to it with such name.}

\begin{figure}[h!]
  \vspace{0.35cm}
  \centering
  \includegraphics[width=0.8\linewidth]{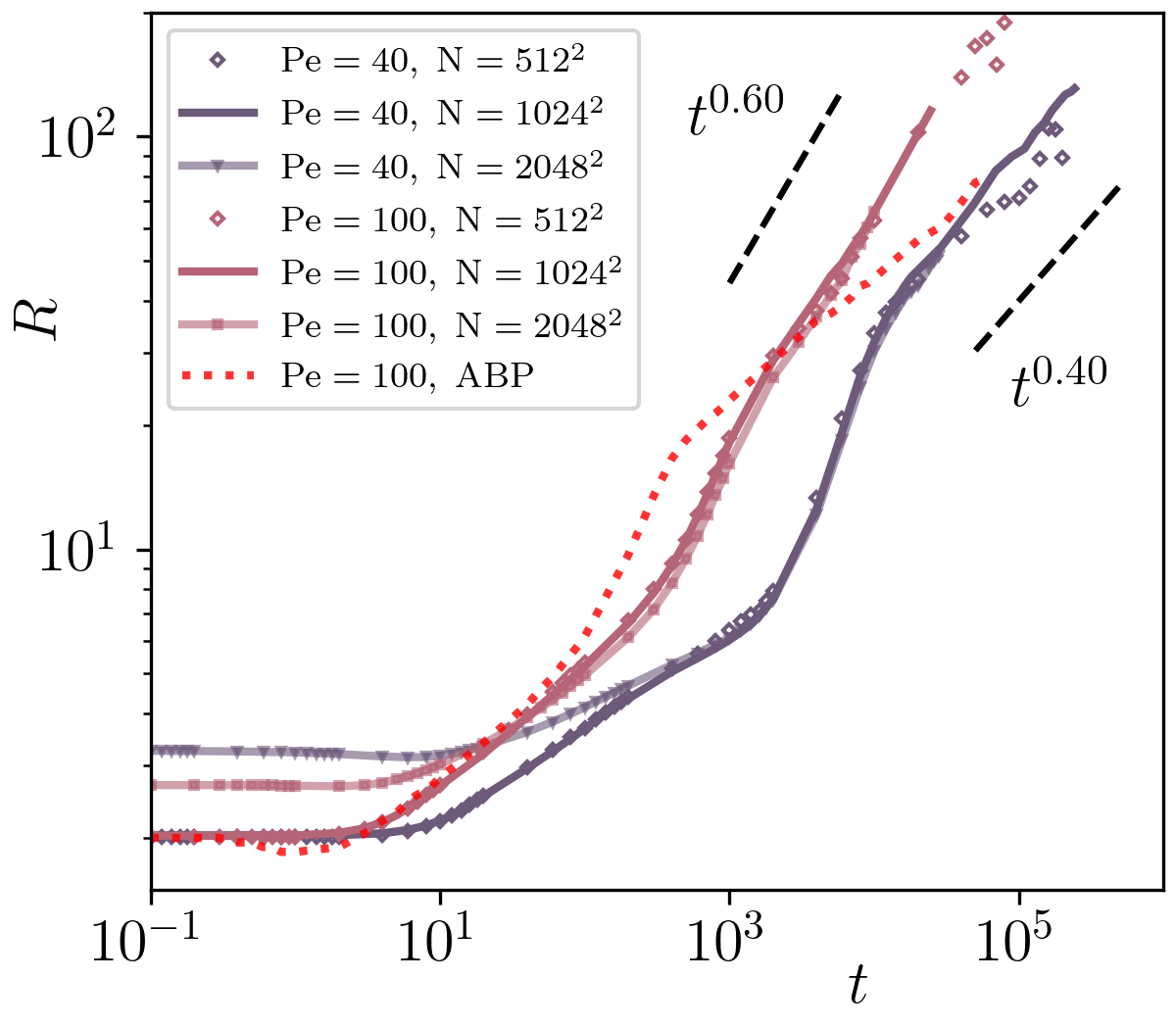}
  \caption{{\bf Finite size dependence of the growing length.} Comparison between dumbbells' data for 
  $N=512^2/2$, $N=1024^2/2$ and $N=2048^2/2$
    at Pe  = 40 (blue solid lines and datapoints below) 
    and Pe = 100 (purple solid lines and datapoints above), and the ABP's growing length 
    at Pe = 100 with $N=1024^2/2$  (dotted red line). {Densities are the same as those considered in Fig.~\ref{fig:Rt}.}
  }
  \label{fig:finite-size scaling}
\end{figure}

{We checked that the curves of $R(t)$  are not affected by significant finite size effects, by considering  larger system sizes of $N= 512^2/2, 1024^2/2$ and $2048^2/2$ dumbbells, see Fig.~\ref{fig:finite-size scaling}. Only at 
$N= 512^2/2$, the data become noisier for the small Pe and deviations from the asymptotic behaviour are visible at long times.} 
At even later times,  the  size  of  the  dense  phase saturates
to  a value  that is proportional to the linear  system  size as will be clear from the analysis of the 
averaged gyration radius of the dense clusters, displayed in Fig.~\ref{fig:saturation}.

For the weakest activity, Pe = 20, the behavior of $R(t)$  is quite different (Fig.~\ref{fig:Rt}(g)). The typical  length 
enters directly the last algebraic growth, without an intermediate rapid growth. 
The difference between the behaviour at weak and strong activities can be 
traced back to the fact that, while the 50:50 curve at high Pe lies at relatively 
low packing fractions, $\phi \sim 0.5$, at low Pe this curve is at very high global densities, $\phi\sim 0.75$, 
and, moreover, the $\phi$ interval over which there is co-existence of dilute and dense phases is very narrow.
The dumbbells are already quite packed initially and the density of the gas is also very high at low Pe.
The global conditions are therefore quite different from the ones at high Pe.  The difference between the high and low Pe 
 behaviors  can be appreciated from the snapshots in Fig.~\ref{fig:Rt}(a)-(c). They not only show the different initial conditions  {(compare Figs.~\ref{fig:Rt}(a) and (d), with average initial density $\phi = 0.50$ and $\phi = 0.70$, respectively)}, but also the fact that at late times (Figs.~\ref{fig:Rt}(c) and (f)) the  density of the gas, determined by the lower limit of the phase-separated 
 region of the phase diagram (Fig.~\ref{fig:phase-diagram})  is much lower at high Pe than at low Pe.


In the last  scaling regime, the typical length calculated from Eq.~(\ref{eq:length}) grows algebraically,
\begin{equation}
  R(t) \sim t^{1/z}
  \; ,
  \label{eq:growing-length}
\end{equation}
with a dynamic exponent $z$ that, for our numerical data, increases with decreasing Pe
from roughly $z = 1.6$ at Pe = 200, to  $z = 2.5$ at Pe = 20. One could expect
the dynamic exponent to approach the standard value of conservative phase
separation, $z=3$, at Pe going to zero, but it is very hard to make reliable measurements at
still lower Pe since the packing fractions with phase separation are very high ($\phi \gtrsim 0.77$, see Fig.~\ref{fig:phase-diagram}).
In all  Pe $\neq 0$ cases that we studied, the growth is faster than for ABPs,
see the red dotted curve also plotted in Fig.~\ref{fig:finite-size scaling},
where we found $z\sim 3$ for all Pe in the MIPS region of the phase diagram~\cite{Caporusso20}.

We recall that the growth of the dense component
of the same active dumbbell model was first studied in Ref.~\cite{gonnella2014phase} .  However, in that paper the system sizes considered were much smaller than the ones we consider here, 
and the separation of time scales (with a late scaling limit) was not established. Growing lengths of the order of the
small box sizes used were quickly reached and only the second regime was observed, with a very fast power law growth with
exponent $1/z \approx 0.9$. Indeed, this large power 
is similar to the one that we measure in the intermediate regime of rapid growth, see the dotted segment in 
Fig.~\ref{fig:Rt}(g). 

\subsection{{Structure factor} }
\label{structfact}

\begin{figure}[t!]
  \centering
  \includegraphics[width=.9\linewidth]{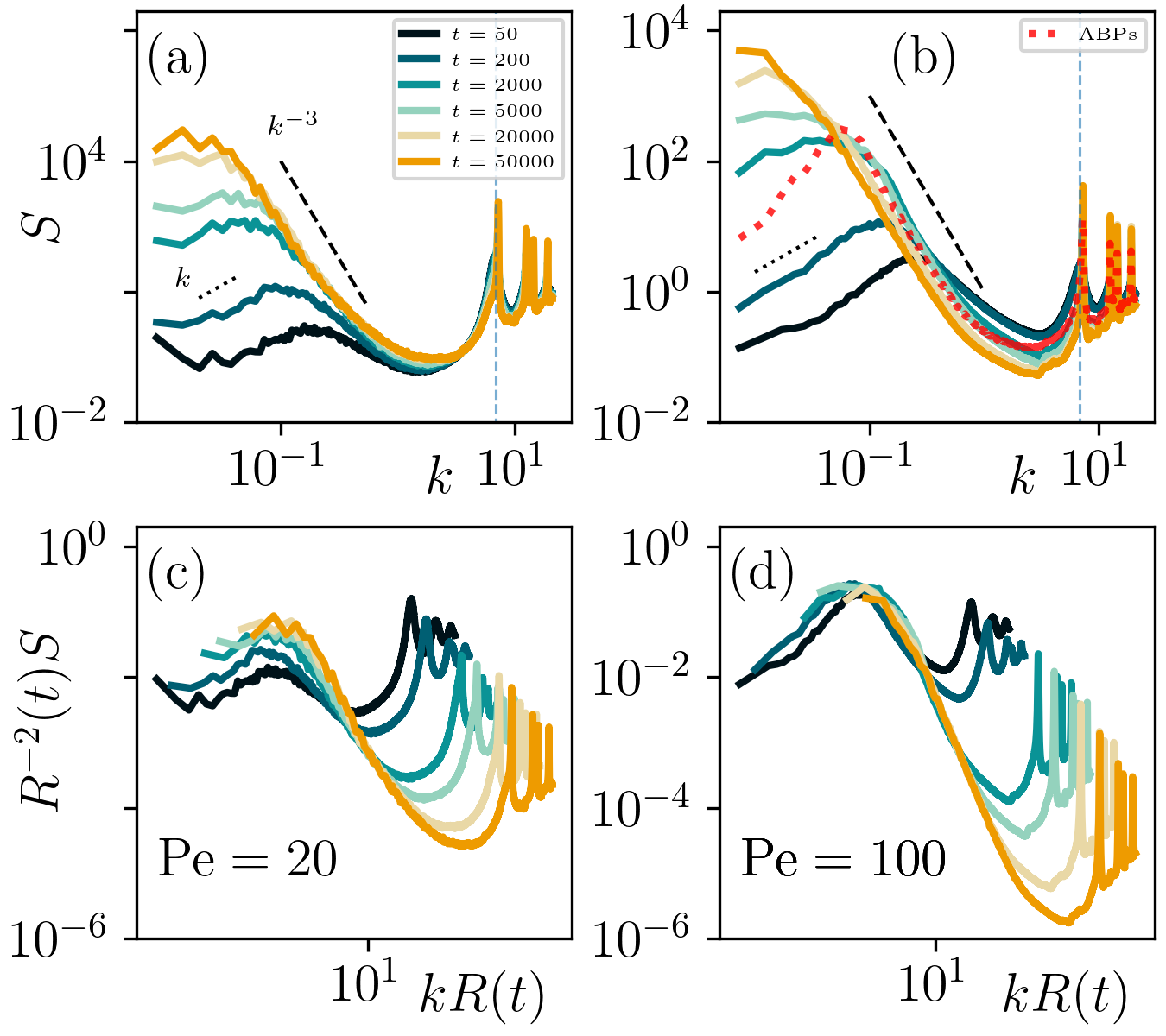}
  \caption{{\bf The dynamic structure factor.}
    In the upper row, the wave vector dependence of the structure factor at different times
    given in the keys,  for Pe = 20 (a) and Pe = 100 (b),
    with densities $\phi = 0.7$ and $\phi = 0.5$, respectively, which locate the system on the 50:50 line in the phase diagram.
    {In the first two panels,} the powers $k^{-3}$ and
    $k$ are indicated with dashed and dotted black lines,
    respectively, {and the vertical blue dotted line represents 
    the time-independent peak related to the crystal structure of the dense phase}. 
    The minimal wave vector due to the finite box is $k_{\rm min} \sim 0.006$, in the
    system with $N=1024^2/2$ used here.   The red dotted curve in (b) corresponds to the  ABPs data. 
    A fit of the Pe = 20 data would yield a weak deviation from Porod's law, $k^{-a}$, with $a \sim 2.6$.
    In the lower row, scaling plots according to
    Eq.~(\ref{eq:scaling}) with $R(t)$ extracted from the numerical data (no fit assumed).
  }
  \label{fig:structure-factor}
\end{figure}

{We now analyze in more detail how the structure factor evolves in time during  growth.}
The structure factors for two extreme values of Pe, Pe = 20 and Pe = 100, and densities of Fig.~\ref{fig:Rt}, are displayed in Fig.~\ref{fig:structure-factor}(a)-(b) at different times, from $t=50$ to $t=5\times 10^4$. {Before commenting the structure factor functions, we need to observe that while for the Pe = 20 case,
at $t\sim 50$ the growth length $R(t)$ has already reached its asymptotic algebraic form (the scaling regime),  for Pe = 100  the growth length is in the rapidly increasing intermediate regime and crosses over to the 
asymptotic algebraic one only at  $t\,$\raisebox{-0.1cm}{$\stackrel{>}{\sim}$}$\, 10^3$
 (Fig.~\ref{fig:Rt}(g)).} 

The structure factors in Fig.~\ref{fig:structure-factor}(a)-(b) have a very clear peak at a time-independent large wave-length, $k \sim 2\pi/\sigma_{\rm d}$
related to the short-distance crystalline structure of the dense phase. We also have the presence of time-independent secondary peaks at multiple frequency values $2\pi n / \sigma_{\rm d}$, with $n$ integer. 
These peaks have a lower intensity than the one at $k \sim 2\pi/\sigma_{\rm d}$,  and are also characteristic of the crystalline  order. 

More interesting is the low $k$ peak which does depend on time, and moves towards smaller values as
time increases. It is related to the progressive growth of the typical cluster size of the dense component.

The wave vector dependence, in between the first and second characteristic values, is algebraic
$S(k, t) \sim k^{-a}$. For segregated systems with sharp and regular interfaces between the phases,
the Porod law yields $a=d+1$. For Pe = 20 we measure $a \sim 2.6$, possibly
due to the fact that the high global density makes the interfaces be contaminated by nearby
dumbbells of the gas.
For Pe = 100, instead, the power is very close to the expected $a=3$, see the dashed line in the figure, 
practically identical to the one measured for ABPs (red dotted curve)~\cite{Caporusso20}.

At strictly $k=0$ the structure factor that we define should equal $2N+1$. However, due to the
finite size box, the minimal wave vector that we can access is $2\pi / L$ and, for the spherically 
averaged $S(k,t)$ a slightly bigger $k_{\rm min}$ guarantees a smoother result. This is the reason 
why the curves in Fig.~\ref{fig:structure-factor} start from a larger value of $k$.


In the scaling regime,  on the left of the first  {low $k$} peak, $k$~\raisebox{-0.1cm}{$\stackrel{<}{\sim}$}~$2\pi / R(t)$, 
{it is easy to spot a linear behavior for Pe = 100 of the structure factor as a function of $k$. For Pe = 20, instead, the structure factor appears flat.}
This is different from what was measured in 
ABP systems~\cite{Caporusso20}, where a $k^2$ behaviour was found, see the dotted red lines on the left in Fig.~\ref{fig:structure-factor}(b). 
The weak $k$ behavior 
is related to the way in which the variance of the number of particles within a spherical shell
$\sigma^2_N$ scales with its radius $r$.  A weaker power would indicate 
less fluctuations, and we could associate this to the fact that the disks, being attached to another 
one to form the dumbbell, allow for less fluctuations in the molecular system.
{It is interesting to notice that a similar linear behaviour in the low-$k$ regime is observed in hyperuniform fluids of active particles \cite{lei2019nonequilibrium}.}



The plots  in Fig.~\ref{fig:structure-factor}(c)-(d) 
demonstrate that the data of Fig.~\ref{fig:structure-factor}(a)-(b) satisfy the dynamic scaling
\begin{equation}
  S(k,t) = R^2(t) \; f(k R(t))
  \label{eq:scaling}
\end{equation}
{in the asymptotic algebraic regime.  Hence, this justifies naming ``scaling regime'' this time sector}, 
with the length $R(t)$ already calculated from Eq.~(\ref{eq:length}) (no fit assumed).
Using the law in Eq.~(\ref{eq:growing-length}) would yield a similarly good scaling.
We conclude  that the scaling properties of $S(k,t)$ in the dumbbells sample are as good as for the ABP system.

\subsection{{Number and size of clusters}}

\begin{figure}[h!]
\vspace{0.3cm}
  \centering
  \includegraphics[width=0.7\linewidth]{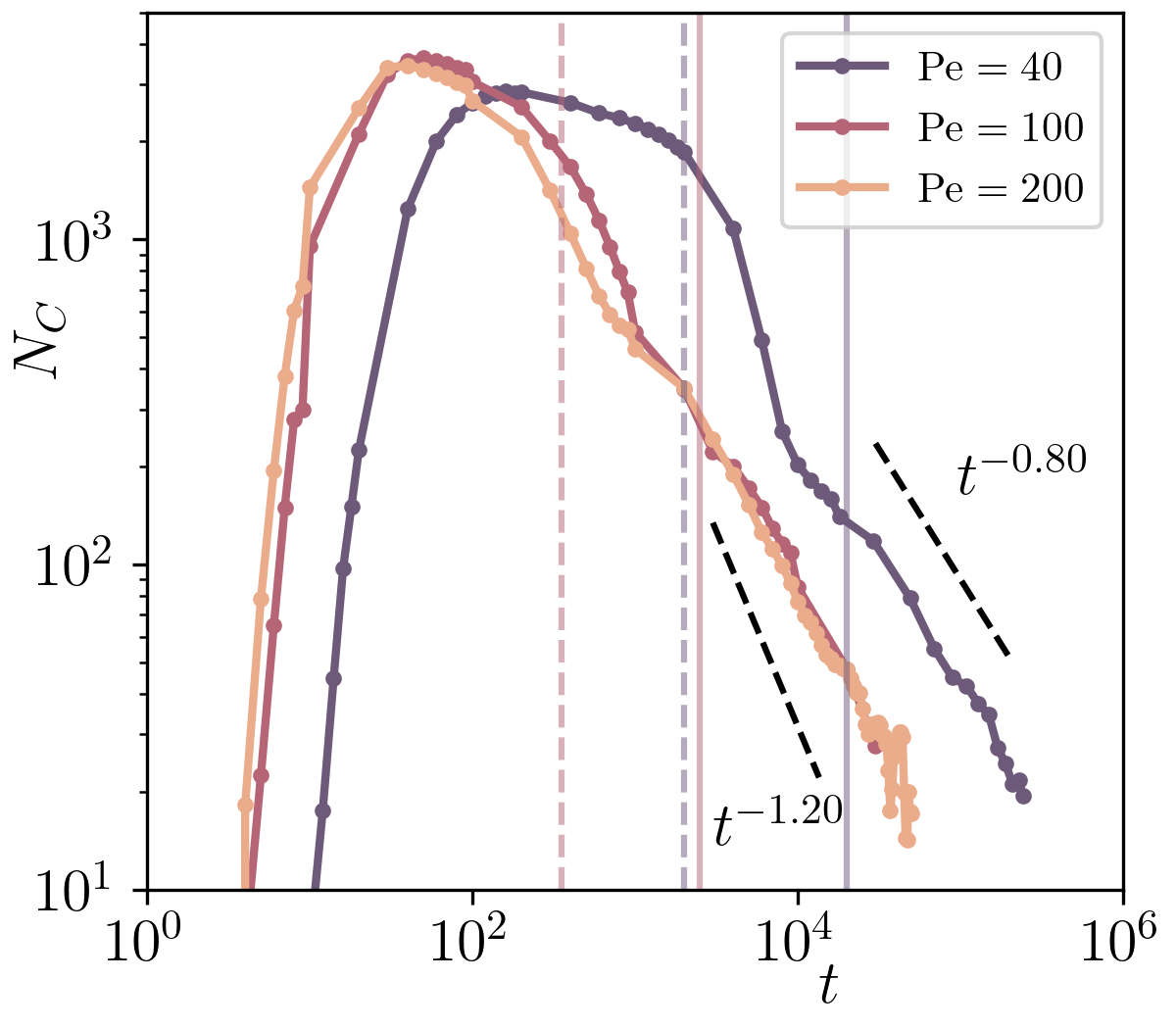}
  \caption{{\bf The number of clusters} as a function of time for three Pe {and the 
  global densities of Fig.~\ref{fig:Rt}}.  The dashed lines
    are the algebraic decays $t^{-2/z}$,  with $z\sim 1.67$ at Pe = 100, 200 and $z\sim 2.5$ for Pe = 40, the
    values extracted from the analysis of the structure factor.
    {The vertical  dashed and continuous lines indicate the transition to the fast growing and the scaling regime, respectively, of the growing length
    for the two Pe = 40 and 200 (grey and orange lines) cases.}
  }
  \label{fig:number-clusters}
\end{figure}

{We now analyze more quantitatively the evolution of the number of clusters for sufficiently
large Pe $\ge 40$, and link it to  the
 three dynamic regimes.} The algorithm to identify clusters has been  explained
in Sec.~\ref{subsec:cluster-identification}. Data are presented in Fig.~\ref{fig:number-clusters}
for Pe = $40, 100$ and 200 and the same densities as in Fig.~\ref{fig:Rt}. {The vertical light lines delimit 
 the three regimes found from the analysis of the growing length.}

In the first rapidly nucleating regime, we observe that the number of clusters $N_C$
grows fast until a maximum {is reached}. {Afterwards, we observe a progressively rapid decay in $N_C$, corresponding to an intermediate algebraic growth phase where droplets grow through condensation and aggregation. When the  scaling regime is reached, the decay achieves an algebraic form with  an exponent $2/z$, }
\begin{equation}
  N_C \sim t^{-2/z},
\end{equation}
that is roughly consistent with the growth of the typical length $R\sim t^{1/z}$.
The global form of these curves is the same as the
one for ABPs, see Fig.~2(b) in~\cite{Digregorio18}. 

\begin{figure}[h!]
\vspace{0.5cm}
  \centering
  \includegraphics[width=1.0\linewidth]{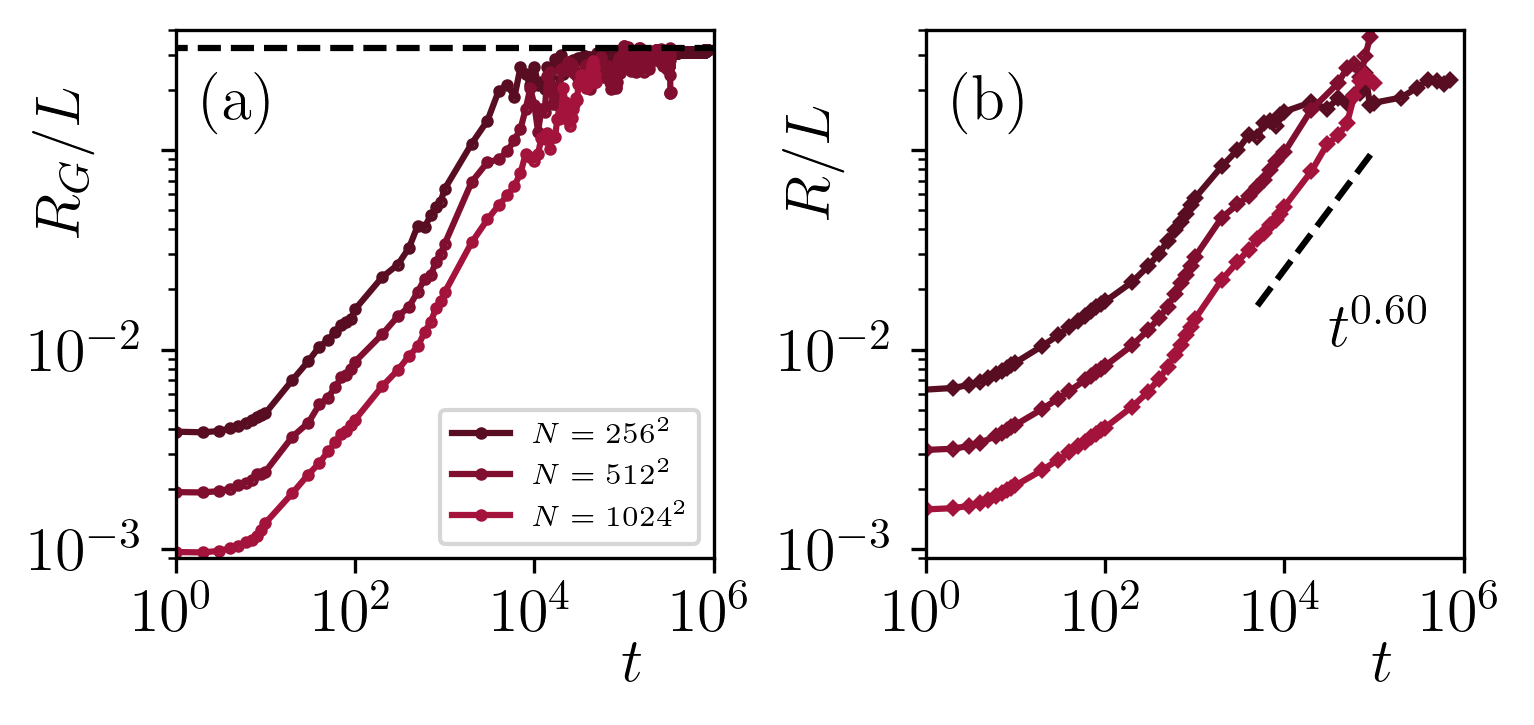}
  \caption{{\bf Finite size dependence and saturation of the growing length.} 
  (a) The averaged gyration radius of the clusters, normalized by the linear system size L, for systems with different
    numbers of dumbbells, all at  Pe = 100 and  same global density of Fig.~\ref{fig:Rt}.
       (b) The growing length of Fig.~\ref{fig:finite-size scaling} normalized by the system size.
     The dashed line in (b)  represents the growth law $t^{0.6}$, and the horizontal one in (a)
    is at $R_G/L \approx 0.33$, 
    and indicates the saturation of the size of the dense component
    with system size.   }
  \label{fig:saturation}
\end{figure}

Having identified the clusters, we can also measure their individual gyration radius as
\begin{equation}
  R^\alpha_G(t) = \left[ \frac{1}{N_\alpha} \sum_{i \in \alpha} ({\mathbf r}_i(t) - {\mathbf r}^\alpha_{\rm cm}(t))^2 \right]^{1/2}
  \!\!\! ,
\end{equation}
with ${\mathbf r}^\alpha_{\rm cm}(t)$ the position of the center of mass of the $\alpha$th cluster and
$N_\alpha$ the number of beads in that cluster, 
and next average it over all $\alpha=1, \dots, N_C$ clusters. This gives us another way to estimate
the typical growing length of the dense component. We display these measurements in Fig.~\ref{fig:saturation} (a),
using different system sizes, all at the Pe = 100 and  packing fraction of Fig~\ref{fig:Rt}. The curves{, scaled by the system size $L$,} increase monotonically in 
time, and have roughly the same behaviour as $R(t)$ measured from the structure factor for the same parameters {(Figs.~\ref{fig:finite-size scaling} and~\ref{fig:saturation}(b), the latter scaled by $L$)}.
At sufficiently long times $R_G$, as well as $R(t)$,  saturates to a size dependent value 
(which needs not be the same because of the slightly different definitions). 
 Indeed, in the very long time limit, the dense phase approaches a
finite fraction of the total system, that only depends on the ratio between the dense and the dilute phase area
and indirectly on Pe and $\phi$, and is represented in Fig.~\ref{fig:saturation}(a)
by the horizontal dashed line. 

\subsection{{Gas phase}}

\begin{figure}[t!]
  \centering
  \includegraphics[width=0.9\linewidth]{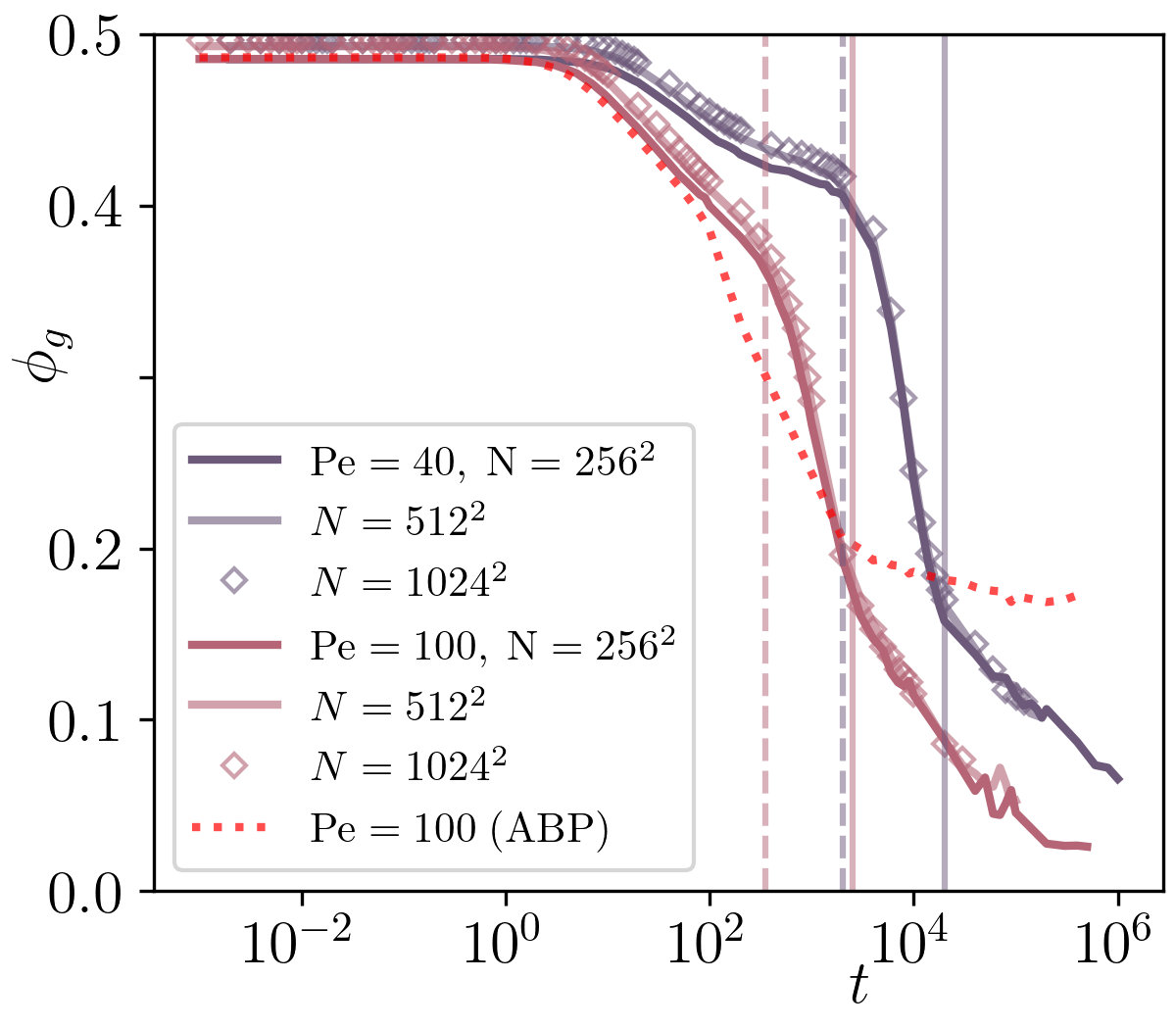}
  \caption{{\bf The surface fraction occupied by the gas}. The lower purple curves are for Pe  = 100, 
  and the upper blue curves for Pe = 40,  in systems with $2N=256^2$, $512^2$ and $1024^2$ particles, and same densities as of Fig.~\ref{fig:Rt}. 
  The dotted red line represents the surface fraction occupied by the gas in the ABPs systems with Pe = 100 and same global density $\phi = 0.5$. 
  {The vertical  dashed and continuous lines indicate the transition to the fast growing and the scaling regime, respectively, of the growing length
    for  the Pe = 40 and 100 (grey and purple lines) cases.}}
  \label{fig:gas}
\end{figure}

Finally, in Fig.~\ref{fig:gas} we show the surface fraction occupied by the gas or dilute phase, $\phi_g$, as a function of time,
for the same Pe values and global packing fractions used in the previous plots. {The vertical light lines are located at the 
limits of the three regimes, as found from the analysis of the growing length.}

The first regime is characterized by a slow decrease of the density of the gas, due to the nucleation of small droplets, 
also suggested by the simultaneous increase of the number of clusters (Fig.~\ref{fig:number-clusters}) 
and in the development of a peak
in the structure factor (Fig.~\ref{fig:structure-factor}).
The second regime is characterized by a fast decrease of the gas density. As could be seen from the movies, 
this regime is characterized by
the coalescence of the droplets, that also increase in size absorbing particles from the gaseous phase.  
The first two regimes have a similar behavior to the one of ABP systems,  shown with a dotted red curve.

We note that, on the contrary with ABPs, there is still a decay of gas density
in the scaling regime. Saturation is just beginning to manifest in the Pe = 100 case, whereas it has not been attained within the simulation times for Pe = 40. 

\section{Orientational order of the dense phase}
\label{sec:orientational}

We now quantify in more detail the internal organization of the dumbbells within the
dense clusters during its growth. We focus here on the analysis of the dumbbells disks orientational order,
as quantified by the local hexatic order parameter
\begin{equation}
  \psi_{6j} = \frac{1}{N_j} \sum_{k\in \partial_j} e^{i6 \theta_{jk}}
  \; ,
  \label{eq:local-hexatic-op}
\end{equation}
with $ \theta_{jk}$ the angle formed  by the segment that connects the center of the $j$th disk and the one of its
$k$th, out of $N_j$, nearest neighbors according to a Voronoi construction, and a reference, say horizontal, direction.

Large dense clusters can show a polycrystalline structure, {\it i.e.} be a mosaic of different hexatically ordered domains.
Indeed, the clusters that  aggregate  do not  necessarily  share  the  same {average direction of} $\psi_{6j}$, and grain boundaries appear in the growing dense phase.
These interfaces may heal and let the orientational order progressively grow or not.
In order to identify these domains and measure their average size, $R_H$,
we identify the hexatic domains according to the argument of $\psi_{6j}$
or by the gradient of its modulus, coarse-grained over a small cell.
The two methods give consistent results.
More precisely, we follow the procedure described
in Ref.~\cite{Caporusso20}, where we redirect the reader for more details.
We discretize the phase of the local hexatic order parameter (\ref{eq:local-hexatic-op})
into $n=6$ bins and we split the system accordingly.
We then apply the DBSCAN algorithm to each part of the system separately.


\subsection{Hexatic growing length}

The analysis of the average  of the hexatic length over all dense clusters, which we call $R_H$, is shown in 
Fig.~\ref{fig:hexatic} (a).
Concomitantly with the growth of the dense phase, we see that the growth of hexatic order
follows the same three {dynamical regimes}. After the early {nucleation} {stage}, the evolution enters the intermediate regime
in which $R_H \sim t^{0.8}$ as for ABPs, see the red dotted curve in Fig.~\ref{fig:hexatic}(a). 
At a smooth crossover  the
evolution slows down {in the scaling regime}, but not as dramatically as for ABPs. Here, we find
\begin{eqnarray}
  R_H \sim
  \left\{
  \begin{array}{ll}
    t^{0.27} \qquad \mbox{low Pe}
    \\
    t^{0.4} \qquad \mbox{high Pe}
  \end{array}
  \right.
\end{eqnarray}
while for ABPs in MIPS the exponent was close to $0.13$~\cite{Caporusso20}. These
features are shown in Fig.~\ref{fig:hexatic} (a) {with dashed lines}. 

{We confirmed (not shown) that in the scaling regime the
dynamic behaviour is not affected if we increase the system size.}

\subsubsection{Dependence on  activity}

\begin{figure}[h!]
  \centering
  \includegraphics[width=0.98\linewidth]{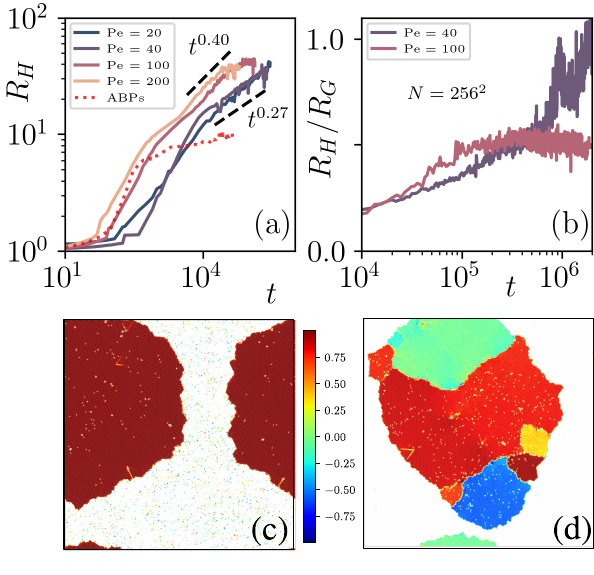}
  \caption{{\bf The hexatic order.} (a) The averaged length of the orientationally ordered parts of the dense component,
    for different Pe and for densities on the 50:50 curve in the phase diagram of Fig.~\ref{fig:phase-diagram}. The red dotted line shows the data for the ABP system.
    (b) Comparison between full orientational ordering at low Pe and saturation to a patchwork
    of different orientational orders at high Pe, averaged over 10 independent runs. In (c) and (d) two examples of the asymptotic configurations
    at low ($\rm{Pe} = 40$) and high ($\rm{Pe} = 100$), respectively. }
  \label{fig:hexatic}
\end{figure}

{Between Pe = 40 and Pe = 100}
we observe a clear change in the behaviour of $R_H$ at long times, in the last 
scaling regime.
For weaker activities, $R_H$ continues to grow towards  the typical size of the dense component, $R_G$, which
 thus acquires  a uniform orientational order. Instead, for stronger activities
$R_H$ approaches a finite limit, $R^s_H$, that, as we argue below, can be identified with a metastable situation. 
Figure~\ref{fig:hexatic} (b)
displays these features: while  the ratio $R_H/R_G$ tends to one for Pe = 40, it remains blocked at a value close to 0.5
for Pe = 100.  

{Figures~\ref{fig:hexatic}(c)-(d) illustrate that full orientational order
is reached at low Pe, while at high Pe we have saturation into patches of different colors, representing hexatically ordered clusters with the {local} hexatic parameter pointing in different directions. }



\subsubsection{Dependence on packing fraction}

\begin{figure}[b!]
\vspace{0.5cm}
  \centering
  \includegraphics[width=\linewidth]{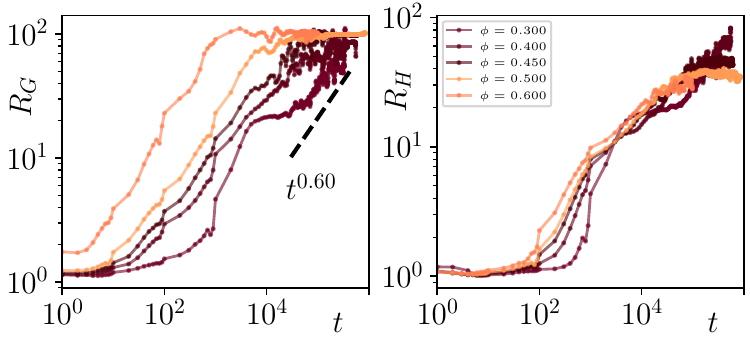}
  \caption{ 
  {\bf Global density effects} on the growth of the dense phase gyration radius $R_G$ and hexatic radius $R_H$ {(left and right panel, respectively)}. The systems are at Pe = 100 and varying packing fractions 
  reported in the legend. The number of particles is fixed to  $2N = 256^2$ and the packing fraction is tuned by changing the 
  size of the simulation box. {Note that for this system size the growing regimes are less visible.}
  }
  \label{fig:density}
\end{figure}

The dependence of the hexatic order on the global packing fraction, at fixed Pe, is analyzed in Fig.~\ref{fig:density}
where $R_G$ and $R_H$ are plotted as a function of time. We reckon that at low $\phi$, even for high Pe, the 
dense phase eventually reaches a unique orientational order, while this is not the case at high $\phi$.

The way in which the sub-domains get blocked or the dense phase manages to order orientationally 
can be appreciated in the movies~M3 and M4.

Figure~\ref{fig:density} gives further  support to the fact that the saturation of $R_H$ reached at high Pe is only a metastable feature. Changing the density,  $R_G$ tends {to a similar asymptotic} 
value, while $R_H$ saturates at {larger values decreasing the density. Thus,} at lower 
global densities dumbbells do not get so much stuck and the hexatic domains can further order reaching larger $R_H$ and 
even the  {asymptotic}  $R_G$ value in some cases.


\subsection{{Stability of hexatic domains in the  dense phase}}

In the analysis shown so far, we found that for low Pe the dense phase reaches a full orientational order
while at high Pe it does not, and the dense phase is made of patches with different hexatic orders. We now 
investigate whether these structures are stable or not. {Note that all the simulations in this section were done using 
a system with $2N = 256^2$.}

First, we took a long-time configuration at Pe = 40, which has completely ordered to the same hexatic parameter,
and we removed the gas. We then restarted the dynamics at a higher value of Pe, 
for which we were not able to reach a full orientational order when we evolved the system from a 
random initial condition. The Pe = 40 initial configuration is stable at all Pe, and the dense phase does not 
break up in differently oriented pieces. 

\begin{figure}[h!]
  \centering
  \includegraphics[width=0.8\linewidth]{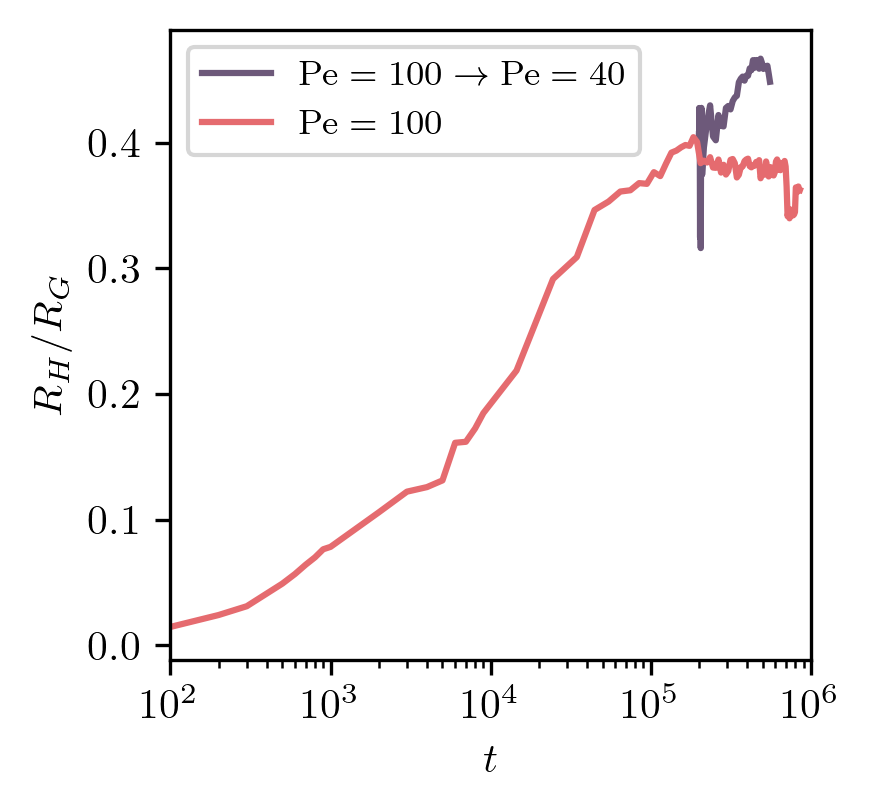}
  \caption{{\bf Quench.} A configuration {with patches of different hexatic orders  at Pe = 100 is instantaneously quenched to Pe = 40 (grey curve), and compared to 
  the same conformation continuing to evolve at Pe = 100 (red curve)}. The ratio between the averaged hexatic radius and the 
  averaged gyration one increases after the quench, implying clusters acquire an homogeneous orientational order at the lower Pe.} 
  \label{fig:stability}
\end{figure}

Conversely, we took a blocked configuration at Pe = 100, with its poly-orientational structure, 
and we suddenly changed Pe, to  Pe = 40, the parameter with which we  
further evolved the system. Right after the quench we 
observed an increase in the averaged hexatic radius, Fig.~\ref{fig:stability}. The 
dense phase tends to better order orientationally at the lower Pe.

{We also proved that} for random initial conditions evolved at even higher Pe, 
systems with sufficiently low density, manage to order and reach larger averaged 
hexatic radii{, compatible with} the averaged radius of gyration itself, 
see Fig.~\ref{fig:density}.

These tests suggest that the truly  stable configurations are the ones with a single orientational order, at all Pe.
The patchworks that we find in the evolution of disordered states at high Pe and high densities would then be metastable
due to the fact that clusters with different hexatic order that collide get stuck and 
the simulations are not long enough to erase the interfaces.

{Note that the behavior of ABP systems is different from the dumbbells' one. For active disk systems, 
if one  starts the dynamics in the MIPS region of the phase diagram from a completely ordered configuration, the dynamics break the hexatic order and 
the hexatic length $R_H$ evolves towards the same value obtained using disordered initial conditions~\cite{Caporusso20}. This  $R_H$  value is smaller than $R_G$ in the asymptotic steady state, with multiple hexatically ordered patches coexisting. 
For this reason, the authors of~\cite{Caporusso20}
claimed that there is a truly stable micro-phase separation of different orientational orders in ABP systems.}
 
{In the case of the dumbbells, in contrast, 
for orientationally ordered initial configurations of the dense phase, 
the cluster does not break over time into smaller patches of different hexatic orders. Conversely, a state with patches of different hexatic orders should eventually be replaced by a fully orientationally ordered cluster.
Moreover, it is important to note that the low Pe behavior found here has no counterpart in the ABP system, since in the latter case the MIPS phase has an ending critical point at a finite Pe.}

\section{Clusters geometry and dynamics}
\label{sec:clusters}

We now turn to the analysis of the clusters' geometry and dynamics, using the tracking algorithm developed in Ref.~\cite{Caporusso22}.
We first investigate the
cluster's geometry in the bulk. Then we characterize the total active force and torque acting on the
clusters. We then identify the typical cluster motion. Adapting the mechanical model derived in~\cite{Caporusso23}
for the motion of 3D dumbbell clusters in dilute conditions, we describe the dynamics of the 2D ones 
analytically.  Of precious help  is the extraction 
of clusters from the bulk of the system and the analysis of their dynamics under isolated conditions.

\subsection{Geometry}
\label{subsec:geometry}

Contrary to ABP clusters, dumbbell ones do not have bubbles within.
We do not see large shape fluctuations either. In sum, they are more stable objects and
the dynamics does not create holes within them, nor lets pieces easily detach,
as one can see happening in ABP clusters~\cite{Caporusso23}.

\begin{figure}[h!]
  \vspace{0.25cm}
  \centering
  \includegraphics[width=0.8\linewidth]{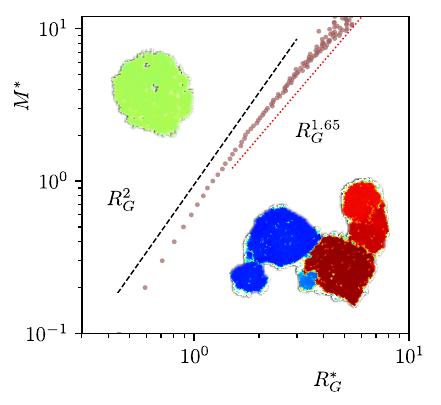}
  \vspace{-0.5cm}
  \caption{{\bf Clusters geometry.} Scatter plot of $M^*=M/\overline{M}(t)$ against $R_G^* = R_G / \overline{R}_G(t)$ {, considering clusters in configurations taken in the scaling regime ($t=1000$) for Pe=100 and density of Fig.~\ref{fig:Rt}. $\overline{M}(t)$ and $\overline{R}_G(t)$ are the average mass and radius of gyration at a given time $t$ considered}. 
  The (black) dashed and (red) dotted lines close to the data indicate the estimates of the fractal dimension 
  of the clusters which are smaller, $d_{\rm f}=2$, and larger, $d_{\rm f}=1.65$, 
  than the average $\overline{R}_G(t)$
  with mass $\overline{M}(t)$. {Typical snapshots of compact (top left) and fractal (bottom right) clusters are displayed.}
  }
  \label{fig:clusters-geometry}
\end{figure}

In Fig.~\ref{fig:clusters-geometry}, we plot the individual cluster mass against its radius of gyration{, considering configurations taken in the scaling regime for Pe=100 at the density of Fig.~\ref{fig:Rt}.}
{Both the mass and radius of gyration are 
normalized by the average values obtained over all clusters at a given time t considered, $M^* = M/\overline M(t)$ and 
similarly for $R_G^*$.}

Similar to our findings for ABP clusters, the scatter plots reveal two regimes: a small mass - short radius of gyration regime in which the two quantities are related by the compact $M^*\sim {R^*_G}^2$ law, and a large mass - long radius of gyration regime with fractal scaling, $M^*\sim {R^*_G}^{d_{\rm f}}$, where $d_{\rm f}\sim 1.65 \neq d =2$. 

This behaviour is qualitatively but also quantitatively similar to the one 
of ABP clusters. The crossover at the average values from compact to 
fractal was also found for ABPs and, moreover, the fractal dimension of the 
large dumbbell clusters is very close to the one of large ABP clusters.

\subsection{Active force and torque}

{In order to understand and quantify the motion of the clusters, two important quantities to consider are the total active force and the active torque} acting on each cluster that we label with $\alpha=1, \dots, N_C$.
The total instantaneous active force is defined as
\begin{equation}
  {\mathbf F}_{\rm act}^\alpha  = \sum_{i\in \alpha} {{\mathbf f}_{{\rm act},i}}.
\end{equation}
The total  instantaneous active torque is defined as
\begin{equation}
  {\bm T}_{\rm{act}}^{\alpha}  = \sum_{i\in \alpha} \, ({\mathbf r}_i - {\mathbf r}^\alpha_{\rm cm}) \times {{\mathbf f}_{{\rm act},i}}
\end{equation}
where $({\mathbf r}_i - {\mathbf r}^\alpha_{\rm cm})$ is the distance between the $i$th bead in cluster $\alpha$
and its center of mass. 

\begin{figure}[h!]
  \centering
  \includegraphics[width=0.8\linewidth]{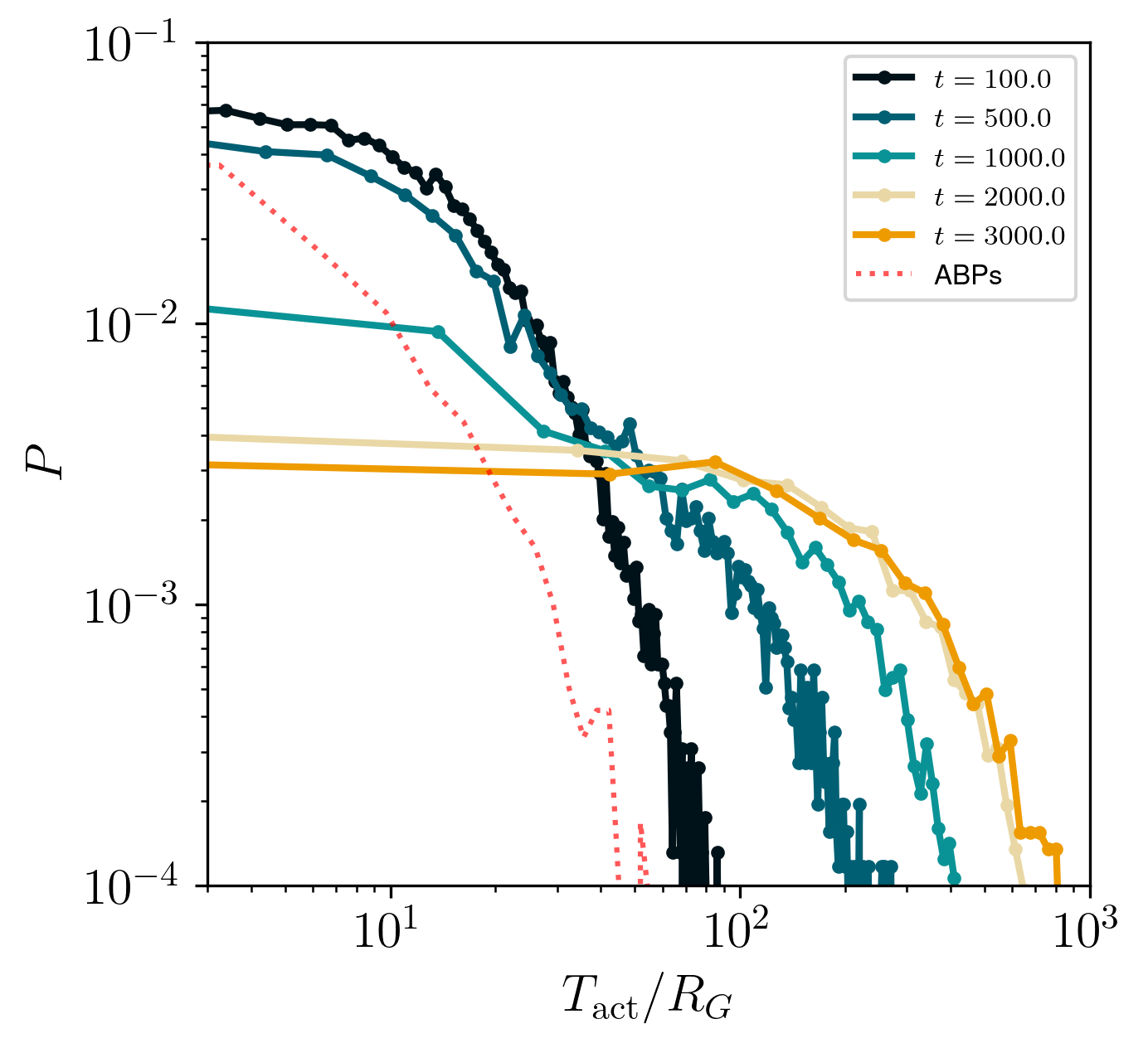}
  \caption{{\bf Distribution of the modulus of the torques} exerted on each cluster, normalized by its gyration radius $R_G$,
  at Pe = 100 {and density of Fig.~\ref{fig:Rt}}, for different instants of the evolution before and during the scaling regime.
  The red dotted curve represents data for the ABPs system in the scaling regime ($t=300$).}
  \label{fig:torque-pdf}
\end{figure}

{Both quantities are typically different from zero and have a  
strong effect on the motion of the clusters, as we will see below.}
The distributions of the modulus of  the torques exerted by the active forces acting on the  clusters 
are shown in Fig.~\ref{fig:torque-pdf} {for Pe = 100 and several times in the rapid growth and in the scaling regimes. While for ABPs we did not see appreciable active torques~\cite{Caporusso22}, for dumbbells} the distributions 
are wide and have weight on quite large values of $T_{\rm{act}}^{\alpha}$. {We observe that these distributions become wider increasing time}. 

\subsection{Motion of clusters in the bulk}

The total active force and torque, which do not vanish, act on the clusters 
and cause a very interesting motion which we now describe. 

\subsubsection{Clusters trajectories}

\begin{figure}[h!]
  \centering
  \vspace{0.25cm}
  \includegraphics[width=\linewidth]{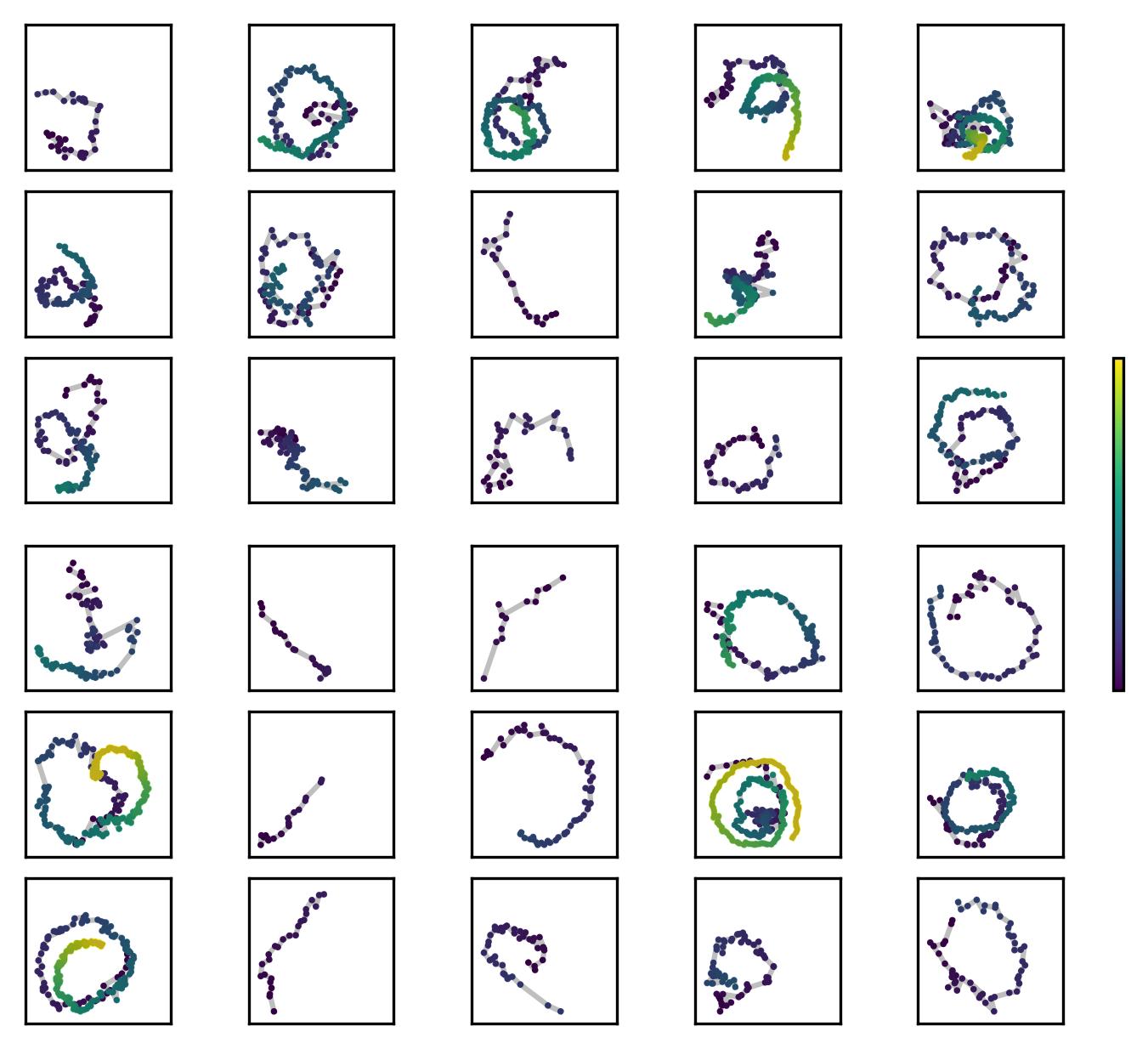}
  \caption{{\bf Typical trajectories of the center of mass of some representative clusters}
  in the interior of a  system with Pe $= 100$, $\phi = 0.35$, $T=0.05${, followed starting from $t_0=1000$ in the regime where clusters have formed and through movement are  colliding with each other. 
  }
 The box size is $10 \sigma \times 10 \sigma$ in all cases. The scale in the right vertical bar represents 
 time, increasing from dark (violet), $t={t_0}$, to light (yellow), $t={t_0+4000}$. The trajectories are colored accordingly. }
  \label{fig:clusters-dynamics}
\end{figure}

Given a cluster {evolving in the whole system (the bulk) with others}, its
center of mass rotates in an approximately circular trajectory 
around a center which {generally does not coincide with the center of mass of the cluster}. 
The cluster behaves as a solid body, with a further rotation around its center of mass. 
The angular velocity around the center of the 
{trajectory} and around its center of mass 
are equal, as we demonstrate numerically below when we extract clusters from the 
bulk.
In Fig.~\ref{fig:clusters-dynamics} we display several of center of mass cluster trajectories. 
The spatial scale is the same in all the panels: $10\sigma_{\rm d} \times 10\sigma_{\rm d}$. 
Some trajectories appear longer, indicating that the selected cluster has moved for an extended period without encountering another cluster, at which point we halt the tracking.
The color scale in the vertical bar on the right 
represents time, evolving from dark (violet) to light (yellow). The shorter trajectories 
look like straight segments but  in practice they are also circular, with 
very large radii, and have been interrupted by collisions with other clusters.

Note that ABD clusters behave very differently in comparison to the motility-induced clusters
of ABP; the latter in fact do not rotate significantly, but only diffuse~\cite{Caporusso22}.

We observe that clusters with a uniform hexatic order and quite 
regular shape are those which turn in {quite regular} spherical trajectories with relatively small radii.
Clusters which resulted from merged smaller clusters with opposite directions of rotation and different hexatic orders separated by interfaces, also follow circular trajectories but with very large radii ({see e.g. movie M5}), and when observed over short time
scales the latter seem to behave ballistically. Over longer time scales, as these clusters are typically quite large, they are modified by collisions with other clusters which changes significantly their kinetics  {(due to the change in the internal dumbbells arrangement, and thus of the total active forces and torques)}.

{Due to clusters collisions and to the exchange of particles between the dense and gas phase, it is difficult to study a trajectory of a single cluster for an extendend period of time. For this reason, in Sec.~\ref{subsec:isolated-clusters} we will extract representative clusters from the bulk and study their 
isolated dynamics over longer time intervals.}

\subsubsection{Mean-square displacement}
\label{subsec:com-msd}

\begin{figure}[h!]
\vspace{0.25cm}
  \centering
  \includegraphics[width=\linewidth]{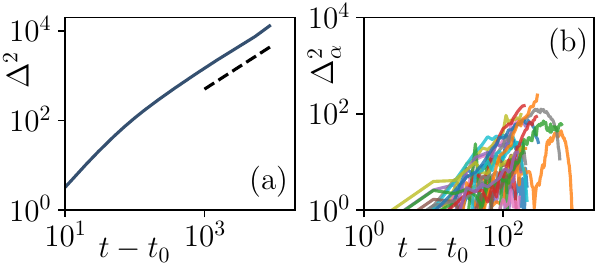}
  \caption{{\bf Averaged mean-square displacement} of (a) the individual dumbbells, with $t_0 = 0$, and (b) the center of mass of the clusters, with $t_0 = 1000$ {(at the beginning of the scaling regime)}. The dotted line represent a linear function. The data are for Pe = 100, $\phi = 0.50$ and $T=0.05$. 
  }
  \label{fig:msd}
\end{figure}

In~\cite{Caporusso23} we used the cluster tracking algorithm to calculate the mean-square 
displacement of the center of mass of each ABP cluster and we found diffusive behavior with a 
mass dependent diffusion coefficient. Consequently, the average over them all also resulted in  diffusion. 
{In the dumbbell case, instead,  clusters have an approximate circular trajectory, and thus cannot diffuse. 
Indeed,  computing the individual mean-square displacement
of the center of mass of each cluster, 
\begin{equation}
\Delta_\alpha^2(t-t_0) = \frac{1}{N_\alpha} \sum_{i=1}^{N_\alpha} \ |{\mathbf r}^{\alpha}_i(t) - {\mathbf r}^\alpha_i(t_0)|^2
\; , 
\end{equation} 
with $\alpha$ labelling the cluster, we see oscillating curves, see 
selected cases  in Fig.~\ref{fig:msd}(b).}

{If one computes the averaged mean-square displacement of {\it all} 
dumbbells, instead, starting from the random initial conformation,}
\begin{equation}
\Delta^2(t-t_0) = \frac{1}{2N} \sum_{i=1}^{2N} \ |{\mathbf r}_i(t) - {\mathbf r}_i(t_0)|^2
\; , 
\end{equation} 
we recover the diffusive behavior, see Fig.~\ref{fig:msd}(a). {Here, the dumbbells in the gas dominate, see the difference in the vertical scales in  Fig.~\ref{fig:msd}(a) and (b).}These claims are also
confirmed by the visual inspection of movies.

\subsection{Motion of isolated clusters}
\label{subsec:isolated-clusters}

In this Section we study the motion of clusters, of regular and fractal kind, {taken from the bulk in the scaling regime, and evolved} in 
isolated conditions.

\subsubsection{Clusters trajectories}

We extract clusters with different characteristics from the 
bulk, similarly to what done in~\cite{Caporusso22}.  We place the clusters in a box without gas, and we evolve them at zero temperature and 
using different values of the active force $f_{\rm act}$, which do not necessarily coincide with the one in the bulk, {and at $T=0$}. Evolving in vacuum, the clusters avoid cluster-cluster collisions and the addition of gas molecules. Under these conditions, 
their mass remains approximately constant for much longer periods of time.
We can therefore follow their dynamics under relatively constant conditions for much longer times than in the bulk. 

\begin{figure}[h!]
\vspace{0.25cm}
  \centering
  \includegraphics[width=\linewidth]{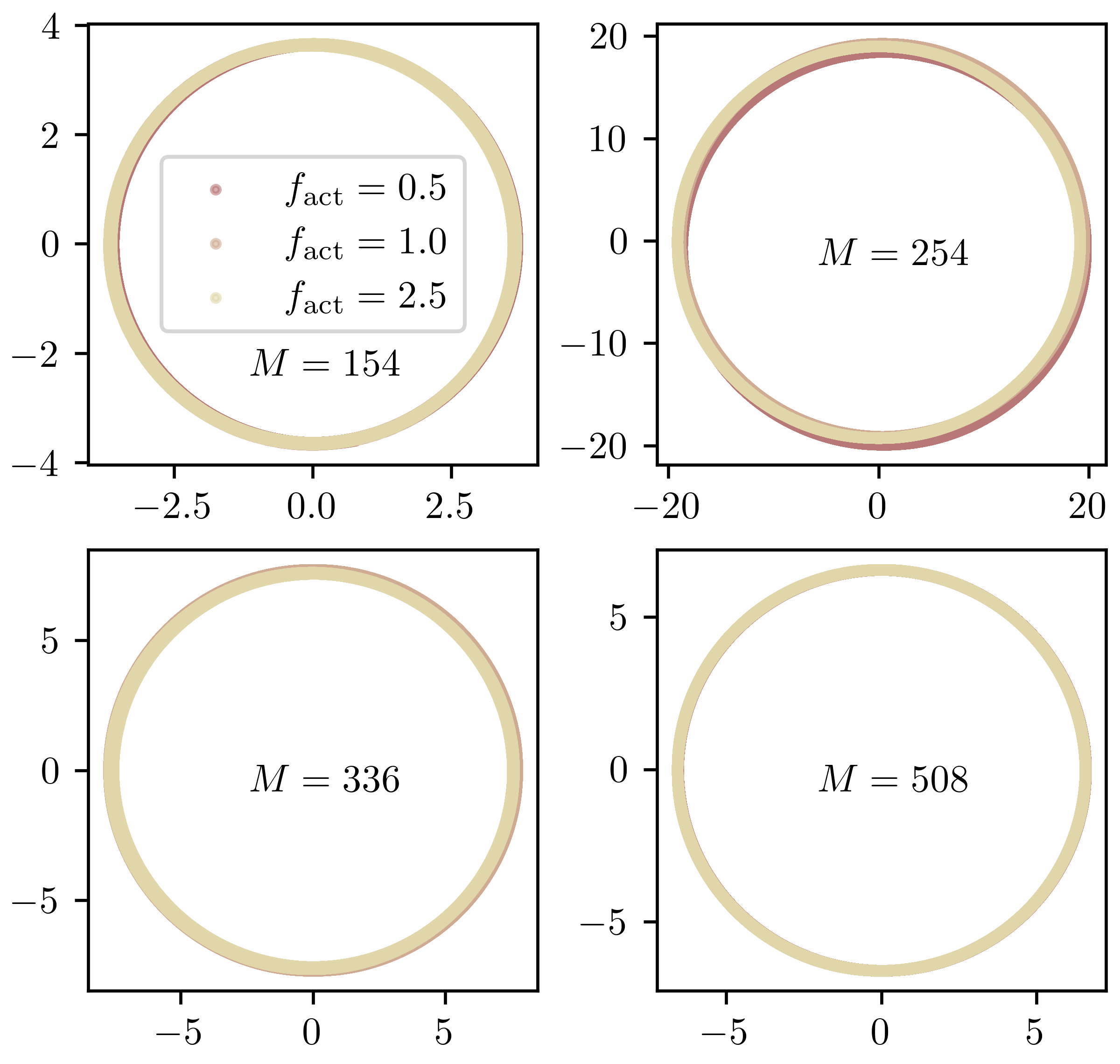}
  \caption{{\bf Isolated clusters with uniform hexatic order.} Motion of the center of mass of 
  four clusters with different mass  
  evolved in vacuum and at zero temperature using different active forces, in the key.
  The spatial scales are 
  not the same in all panels. The trajectories overlap almost perfectly, independently of $f_{\rm act}$.  The values of 
 their mass, radius of gyration, radius of the trajectory
 are given in Table~\ref{table:one}.
    }
  \label{fig:trajectories-vacuum}
\end{figure}

Figure~\ref{fig:trajectories-vacuum} 
shows the motion of four isolated clusters with different initial masses and a single hexatic order, 
evolved using active force values
given in the keys. Notably, the trajectories are quite  independent of the value of $f_{\rm act}$, {and depends only on the geometrical arrangement of the dumbbells inside the clusters}. For the clusters in the 
figure the mass/radius/total active force and total torque relations are given in Table~\ref{table:one}.

\begin{table}[h!]
\begin{tabular}{ccccc}
$M$ & \quad $R_G$  & \quad $R$  & \quad $F_{\rm act}$ & \quad $T_{\rm act}$ 
\\
\hline
\noalign{\vskip 1mm} 
154 & \quad 4.68 & \quad 3.64 & \quad 17.34& \quad 125.86 
\\
254 & \quad 6.03 & \quad 19.14 & \quad 39.13& \quad \ 87.42 
\\
336 & \quad 6.63 & \quad 7.54 & \quad 51.68 & \quad 355.68 
\\
508 & \quad 8.19 & \quad 6.57 & \quad 50.13 & \quad 590.72 
\end{tabular}
\caption{{\bf Features of some representative clusters.} Here are reported the mass, radius of gyration, radius of the circular trajectory, total active force and torque of four different clusters isolated from the bulk { at $f_{\rm act}$=2.5.}
All the quantities in the table are very close to being constant, because these clusters were 
already well stabilized before starting the measurements, and without noise they do not change their 
internal structure. 
The center of mass trajectories are shown in Fig.~\ref{fig:trajectories-vacuum}. {As shown in that figure, changing $f_{\rm act}$, and thus the total active force and torque, does not affect the other quantities.}
}
\label{table:one}
\end{table}

The angular velocity of the center of mass with respect to the center of its circular trajectory
and the angular velocity of a bead with respect to the center of mass of the cluster are equal. This is proven 
numerically by the trajectory of representative clusters, as the one shown in movie~{M6}.

\begin{figure*}
  \centering
  \includegraphics[width=\linewidth]{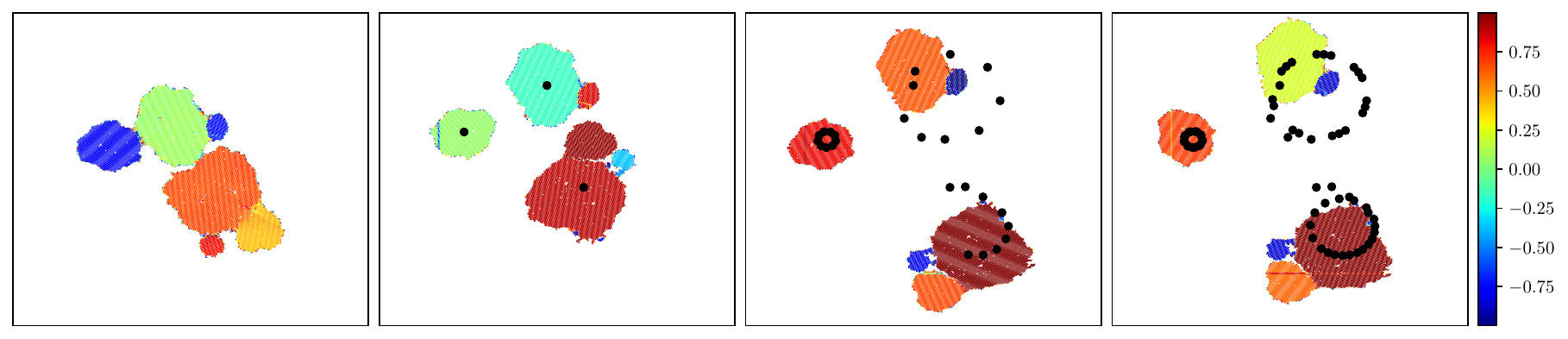}
  \caption{{\bf Isolated clusters with multiple hexatic domains.}  
  Four snapshots of the evolution, in isolation, at time $t=0$, $8000$, $16000$ and $30000$ of a single cluster with heterogeneous orientational order.
  The single cluster breaks  into pieces along the interfaces between the different orientational domains.
  Each of the new clusters, with homogeneous orientational order,  then rotates. The  
  dots traced by their centers of mass are drawn in black. 
  }
  \label{fig:trajectories-heterogeneous-vacuum}
\end{figure*}

Very different is the fate of large clusters with several hexatic orders. They are not stable, they break along the internal 
interfaces and each of the pieces undergoes a noisy circular motion. Their trajectories can be such that they meet again, 
remain attached for some time, break once more and so on and so forth. The snapshots in Fig.~\ref{fig:trajectories-heterogeneous-vacuum} and the
movie~M7
demonstrate these features. This gives support to the fact that stable clusters, although difficult to form, will 
eventually have uniform orientational order.

\subsubsection{{Kinetic model of cluster motion}}
\label{subsec:kinetic-model}

In this section, we derive the equation of motion for a single cluster in isolation, following the approach of \cite{Caporusso23}.

We assume that the cluster is a rigid body, meaning that its shape  {(the internal arrangement of dumbbells)} does not change over time. This is a reasonable approximation for dumbbell clusters that are strongly bound by the active force.

The motion of the center of mass (CoM) of the cluster is described by Newton's second law,
\begin{equation} 
\label{eq:com}
  M \ddot{\bm{R}}_{\text{cm}} = \bm{F}_{\text{act}} + \bm{F}_{\text{drag}}
  \; ,
\end{equation} 
where $2N_c$ is the total number of disks in the cluster,
$M = 2N_c m_{\rm d}$ is its total mass, $\bm{F}_{\text{act}} = \sum_{i=1}^{2N_c} {\mathbf f}_{\text{act},i}$ 
is the total active force acting on the cluster, and 
$\bm{F}_{\text{drag}} = - (M/m_{\rm d}) \gamma_{\rm d} \bm{\dot{R}}_{\text{cm}}$ 
is the total drag force acting on the cluster. 
The internal forces arising from the pair potential cancel out in the equation of motion of the CoM. 

\begin{figure}
  \centering
  \includegraphics[width=.6\linewidth]{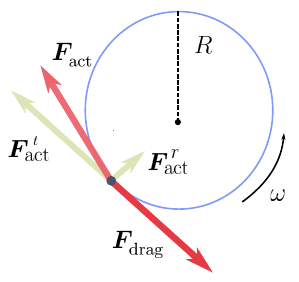}
  \caption{{\bf Sketch of the forces} acting on the center of mass of a cluster
  undergoing circular motion with radius $R$. 
    }
  \label{fig:sketch-forces}
\end{figure}

The forces are drawn in the sketch in Fig.~\ref{fig:sketch-forces}. 
The drag force is tangential to the circular trajectory of the CoM. The total active force 
has a (large) component along  this same tangential 
direction and a (small) component along the radial one.

The equation of motion of the cluster is completed by the equation of motion of the angular momentum, 
\begin{equation}
  \dot{\bm{L}} = \bm{T}_{\text{act}} + \bm{T}_{\text{drag}}
  \; ,
  \label{eq:ang}
\end{equation}
where $\bm{L} = \sum_{i=1}^{2N_c} \bm{r'}_i \times m_{\rm d}{\dot{\bm{r}}}'_i $ is the total angular momentum, $\bm{T}_{\text{act}} = \sum_{i=1}^{2N_c} \bm{r'}_i \times \bm{f}_{\text{act},i}$  is the total active torque acting on the cluster, $\bm{T}_{\text{drag}} = \sum_{i=1}^{2N_c} \bm{r'}_i \times (-\gamma_{\rm d}{\dot{\bm{r}}}'_i)  = - \gamma_{\rm d} \bm{L}/m_{\rm d}$ 
is the total drag torque acting on the cluster, 
and $\bm{r'}_i = \bm{r}_i - \bm{R}_{\text{cm}}$ is the position of the $i$-th particle with respect to the cluster's 
CoM.

We now analyze the  motion of the cluster. 
In an overdamped approximation of Eq.~\eqref{eq:ang}
one can set the time derivative to zero.  
This yields a relation between the modulus of the total active torque and the one of the angular momentum:
\begin{equation}\label{eq:Lomega}
L = \frac{m_{\rm d} T_{\text{act}}}{\gamma_{\rm d}} 
\; . 
\end{equation}

Instead, the overdamped approximation cannot be applied to Eq.~\eqref{eq:com}, since an 
inertial contribution is needed at all times in order to turn the direction of motion of the cluster, as we observe numerically. 
We  thus decomposed the forces in Eq.~\eqref{eq:com}
in radial $\bm{F}^r$ and tangential $\bm{F}^t$ components. 
The former contributes to the change in direction, while the latter to the directed motion only. 
A good approximation~\cite{Caporusso23} is to consider 
the tangential component of the active force  almost fully counterbalanced by the drag force $F^{t}_{\text{act}} \approx \frac{M}{m_{\rm d}} \gamma_{\rm d} \dot{R}_{\text{cm}}$, setting in this direction the time derivative to zero in Eq.~\eqref{eq:com}. 
The remaining small mismatch gives rise to a force
\begin{equation}
\bm{F}^r_{\text{act}} = \bm{F}_{\text{act}} + \bm{F}_{\text{drag}}
\end{equation} 
equal to the inertial contribution of the motion and directed radially towards the center of the trajectory. In this direction, we have $M\ddot R_{\text{cm}}= F^r_{\text{act}}$.

Using the fact that the motion occurs with uniform angular 
velocity $\omega$, we can substitute $\dot{R}_{\text{cm}} = \omega R$ and $\ddot{R}_{\text{cm}} = \omega^2 R$, where $R$ is the radius of the circle (not to be confused with the growing length of previous sections). Thus, 
\begin{align} 
\label{eq:tan}
 & F^{t}_{\text{act}}  = \frac{M}{m_{\rm d}} \gamma_{\rm d}   \omega R \; , \\
 & F^{r}_{\text{act}}  = M \omega^2 R \; . \label{eq:rad}
\end{align}

From Eq.~\eqref{eq:tan}, we can express the radius $R$ of the trajectory in terms of the tangential component of the active force and the angular velocity:
\begin{equation} \label{eq:R}
   R = \frac{m_{\rm d}}{M} \frac{F^{t}_{\text{act}}}{\gamma_{\rm d} \omega}
   \; .
\end{equation}

Finally, expressing the total angular momentum in terms of its moment of intertia, $L=I\omega$, and the latter in terms of the radius of gyration $R_G$ and the mass $M$ as $I = M R_G^2$, using Eq.~(\ref{eq:Lomega}) we can obtain an alternative expression for the radius $R$ in terms of the active force and torque:
\begin{equation} \label{eq:R2}
  R = R_G^2 \, \frac{F^{t}_{\text{act}}}{T_{\text{act}}}
  \; . 
\end{equation} 

{Notably, because both $F^{t}_{\text{act}}$ and $T_{\text{act}}$ are proportional to $f_{\text{act}}$, the model's prediction is that the radius of the trajectory R is independent of $f_{\text{act}}$. Indeed, this result has been already verified numerically in Fig.~\ref{fig:trajectories-vacuum}.}

\begin{figure}
  \centering
  \includegraphics[width=0.8\linewidth]{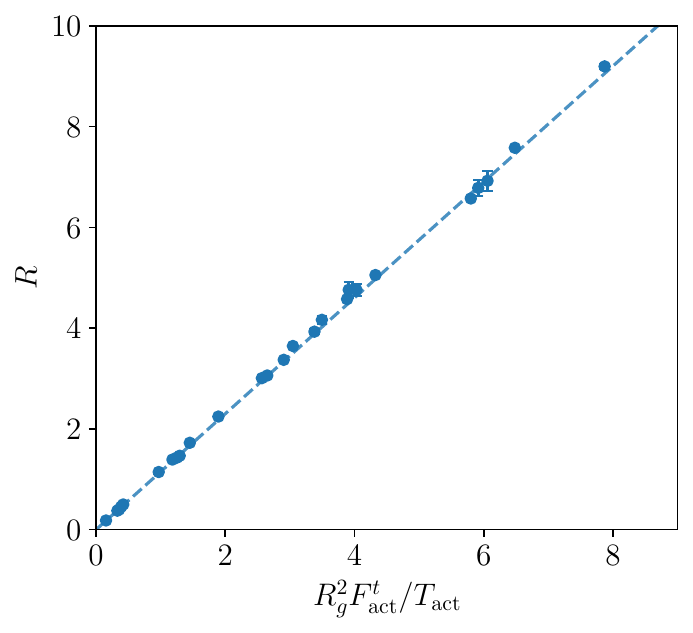}
  \caption{{\bf Radius of the trajectory of the cluster.} The radius of the trajectory of the cluster is measured directly from the clusters' trajectories (blue dots) and compared against the {linear} prevision of Eq.~\eqref{eq:R2} 
  (dashed line).  {The fit 
  yields a slope equal to 1.15.}
   }
  \label{fig:results}
\end{figure}

We tested the validity of Eq.~\eqref{eq:R2} by independently measuring the radius of the trajectory of the cluster, its radius of gyration, and the total active force and torque acting on it. Some concrete values are listed in Table~\ref{table:one}.
The results are shown in Fig.~\ref{fig:results} and are in good agreement with the theoretical prediction{, with the slope of the linear fit $(1.15)$ slightly larger than the expected value of~$1$. This discrepancy indicates that probably there are some higher order corrections to Eq.~\eqref{eq:R2} that are not accounted for by our model.}

\section{Conclusions}
\label{sec:conclusions}

In this paper we studied 
the dynamics of active dumbbell systems, for parameters such that they phase 
separate into dense and dilute components, focusing on the formation of the dense structures.

Our first goal was to characterize the growth of the dense phase from the study of the structure factor. 
We used sufficiently large system sizes, 
with as many as $2048^2/2$ dumbbells,  so as to eliminate any finite size effect during growth. 
For Pe \raisebox{-0.1cm}{$\stackrel{>}{\sim}$} 40 we identified three distinct regimes, similar to the ones observed for ABPs, 
with different time-dependencies of the growing typical length: 
short times transient, rapid growth, and a scaling limit with algebraic growth.
For Pe  \raisebox{-0.1cm}{$\stackrel{<}{\sim}$} 40
        we do not see the intermediate regime, typical of nucleation, and the dynamics seems to enter directly the asymptotic
        scaling limit.
In the scaling regime, the typical length grows algebraically but, contrary to what was found for ABPs~\cite{Caporusso20}, the exponent increases with Pe 
(at fixed percentage of dense and dilute regions), {reaching the value $0.60$ at the highest activity considered}. Moreover,
the growth is considerably faster than for disks.
The values and parameter dependence of this growing length are consistent with 
the ones derived from the decay of the instantaneous number of 
clusters and the averaged gyration radius of the clusters.

Next, we analysed the internal structure of the dense regions paying 
special attention to the orientational order of the disks forming the dumbbells in terms of  
their hexatic arrangement. In the scaling regime, 
the length associated to the orientational  order grows algebraically and
faster than for ABPs. Furthermore, the growth gets faster for larger Pe value,
 with $R_H(t) \sim t^{0.3}$ at low Pe and $R_H(t) \sim t^{0.4}$
        at stronger activities.

An important difference is found at around 
Pe $\sim$ 40: at weaker activity the clusters manage to reach a uniform orientational 
order, while at higher activity this is not achieved and a polycrystalline structure
is formed and gets blocked. The latter behavior is similar to the one found in the MIPS phase of ABPs. 
Quenches between two representative Pe values, and the evolution of fully ordered configurations, 
suggest that for dumbbell systems the truly stable configurations 
are the ones with full orientational order, though these are not accessible dynamically at the
time scales that we can use in the simulations.

A salient feature of the dense ABP clusters is the fact that bubbles pop up at the interfaces between 
different orientationally ordered patches. No such gas bubbles appear in the ABD clusters, not even when there are 
patches of different hexatic order. There are no large shape fluctuations either, with no piece detachment nor any other 
abrupt event. In sum, the ABD clusters are more stable objects 
than the ABP ones. 

Finally, we focused on the dynamics of the clusters, and we performed a tracking 
analysis as the one in~\cite{Caporusso22} for ABPs. In order to perform this study, 
we needed to work at sufficiently high Pe,  Pe \raisebox{-0.1cm}{$\stackrel{>}{\sim}$} 20, otherwise the 
phase separated region of the phase diagram is located at too high global packing fraction
and there are no isolated clusters under these conditions.

Concerning the geometrical properties of the clusters, small clusters are regular while large ones are fractal, with the crossover determined by the 
time dependent growing length, as found for ABP clusters. The fractal dimension measured is $d_{\rm f} \sim 1.65$.

Interestingly enough, the ABD clusters are not diffusive but they rotate as approximate solid bodies
along circular trajectories perturbed by encounters with other clusters and noise.
  

 Extracting clusters from the bulk and following their dynamics in isolation we further proved that large 
 polycrystalline clusters are not stable and break up in pieces of uniform orientational order which then follow their 
 own circular motion.
 
 We adapted the model developed in~\cite{Caporusso23} to describe analytically the motion of the stable 2D clusters. 
 We found very good agreement between the quantitative predictions of the model for the radius 
 of the center of mass circular trajectory of not too large and homogeneous clusters, 
  and the numerical measurements. The concrete prediction is given in Eq.~\eqref{eq:R2} and relates the 
  {radius of the trajectory to the radius of gyration of the cluster itself in a non-trivial manner. Importantly, we found that the radius of the trajectory is independent of the Pe value.}

 {To conclude, the results reported here show the richness of behaviour of the dumbbell model, which differ considerably from the ABP model, and the general importance of considering anisotropic particles
 . These differences span the growth properties,  the predicted steady-state, and the cluster movement, and arise due to the difference in particles interlocking when forming a dense phase.}

\acknowledgements
We acknowledge funding from MIUR Projects No. PRIN
2020/PFCXPE 2022/HNW5YL and ANR-20-CE30-0031. We thank L. Carenza, D. Levis and G. Negro
for very useful discussions.

\bibliographystyle{apsrev4-1}
\bibliography{dumbbellsbiblio.bib}

\begin{thebibliography}{79}%
\makeatletter
\providecommand \@ifxundefined [1]{%
 \@ifx{#1\undefined}
}%
\providecommand \@ifnum [1]{%
 \ifnum #1\expandafter \@firstoftwo
 \else \expandafter \@secondoftwo
 \fi
}%
\providecommand \@ifx [1]{%
 \ifx #1\expandafter \@firstoftwo
 \else \expandafter \@secondoftwo
 \fi
}%
\providecommand \natexlab [1]{#1}%
\providecommand \enquote  [1]{``#1''}%
\providecommand \bibnamefont  [1]{#1}%
\providecommand \bibfnamefont [1]{#1}%
\providecommand \citenamefont [1]{#1}%
\providecommand \href@noop [0]{\@secondoftwo}%
\providecommand \href [0]{\begingroup \@sanitize@url \@href}%
\providecommand \@href[1]{\@@startlink{#1}\@@href}%
\providecommand \@@href[1]{\endgroup#1\@@endlink}%
\providecommand \@sanitize@url [0]{\catcode `\\12\catcode `\$12\catcode `\&12\catcode `\#12\catcode `\^12\catcode `\_12\catcode `\%12\relax}%
\providecommand \@@startlink[1]{}%
\providecommand \@@endlink[0]{}%
\providecommand \url  [0]{\begingroup\@sanitize@url \@url }%
\providecommand \@url [1]{\endgroup\@href {#1}{\urlprefix }}%
\providecommand \urlprefix  [0]{URL }%
\providecommand \Eprint [0]{\href }%
\providecommand \doibase [0]{http://dx.doi.org/}%
\providecommand \selectlanguage [0]{\@gobble}%
\providecommand \bibinfo  [0]{\@secondoftwo}%
\providecommand \bibfield  [0]{\@secondoftwo}%
\providecommand \translation [1]{[#1]}%
\providecommand \BibitemOpen [0]{}%
\providecommand \bibitemStop [0]{}%
\providecommand \bibitemNoStop [0]{.\EOS\space}%
\providecommand \EOS [0]{\spacefactor3000\relax}%
\providecommand \BibitemShut  [1]{\csname bibitem#1\endcsname}%
\let\auto@bib@innerbib\@empty
\bibitem [{\citenamefont {Bechinger}\ \emph {et~al.}(2016)\citenamefont {Bechinger}, \citenamefont {Leonardo}, \citenamefont {L\"owen}, \citenamefont {Reichhardt}, \citenamefont {Volpe},\ and\ \citenamefont {Volpe}}]{bechinger2016}%
  \BibitemOpen
  \bibfield  {author} {\bibinfo {author} {\bibfnamefont {C.}~\bibnamefont {Bechinger}}, \bibinfo {author} {\bibfnamefont {R.~D.}\ \bibnamefont {Leonardo}}, \bibinfo {author} {\bibfnamefont {H.}~\bibnamefont {L\"owen}}, \bibinfo {author} {\bibfnamefont {C.}~\bibnamefont {Reichhardt}}, \bibinfo {author} {\bibfnamefont {G.}~\bibnamefont {Volpe}}, \ and\ \bibinfo {author} {\bibfnamefont {G.}~\bibnamefont {Volpe}},\ }\href@noop {} {\bibfield  {journal} {\bibinfo  {journal} {Rev. Mod. Phys.}\ }\textbf {\bibinfo {volume} {88}},\ \bibinfo {pages} {045006} (\bibinfo {year} {2016})}\BibitemShut {NoStop}%
\bibitem [{\citenamefont {Ramaswamy}(2010)}]{Ramaswamy10}%
  \BibitemOpen
  \bibfield  {author} {\bibinfo {author} {\bibfnamefont {S.}~\bibnamefont {Ramaswamy}},\ }\href@noop {} {\bibfield  {journal} {\bibinfo  {journal} {Ann. Rev. Cond. Matt. Phys.}\ }\textbf {\bibinfo {volume} {1}},\ \bibinfo {pages} {323} (\bibinfo {year} {2010})}\BibitemShut {NoStop}%
\bibitem [{\citenamefont {Vicsek}\ and\ \citenamefont {Zafeiris}(2012)}]{Vicsek12}%
  \BibitemOpen
  \bibfield  {author} {\bibinfo {author} {\bibfnamefont {T.}~\bibnamefont {Vicsek}}\ and\ \bibinfo {author} {\bibfnamefont {A.}~\bibnamefont {Zafeiris}},\ }\href@noop {} {\bibfield  {journal} {\bibinfo  {journal} {Phys. Rep.}\ }\textbf {\bibinfo {volume} {517}},\ \bibinfo {pages} {71} (\bibinfo {year} {2012})}\BibitemShut {NoStop}%
\bibitem [{\citenamefont {Marchetti}\ \emph {et~al.}(2013)\citenamefont {Marchetti}, \citenamefont {Joanny}, \citenamefont {Ramaswamy}, \citenamefont {Liverpool}, \citenamefont {Prost}, \citenamefont {Rao},\ and\ \citenamefont {Simha}}]{Marchetti13}%
  \BibitemOpen
  \bibfield  {author} {\bibinfo {author} {\bibfnamefont {M.~C.}\ \bibnamefont {Marchetti}}, \bibinfo {author} {\bibfnamefont {J.~F.}\ \bibnamefont {Joanny}}, \bibinfo {author} {\bibfnamefont {S.}~\bibnamefont {Ramaswamy}}, \bibinfo {author} {\bibfnamefont {T.~B.}\ \bibnamefont {Liverpool}}, \bibinfo {author} {\bibfnamefont {J.}~\bibnamefont {Prost}}, \bibinfo {author} {\bibfnamefont {M.}~\bibnamefont {Rao}}, \ and\ \bibinfo {author} {\bibfnamefont {R.~A.}\ \bibnamefont {Simha}},\ }\href@noop {} {\bibfield  {journal} {\bibinfo  {journal} {Rev. Mod. Phys.}\ }\textbf {\bibinfo {volume} {85}},\ \bibinfo {pages} {1143} (\bibinfo {year} {2013})}\BibitemShut {NoStop}%
\bibitem [{\citenamefont {Gonnella}\ \emph {et~al.}(2015{\natexlab{a}})\citenamefont {Gonnella}, \citenamefont {Marenduzzo}, \citenamefont {Suma},\ and\ \citenamefont {Tiribocchi}}]{gonn15}%
  \BibitemOpen
  \bibfield  {author} {\bibinfo {author} {\bibfnamefont {G.}~\bibnamefont {Gonnella}}, \bibinfo {author} {\bibfnamefont {D.}~\bibnamefont {Marenduzzo}}, \bibinfo {author} {\bibfnamefont {A.}~\bibnamefont {Suma}}, \ and\ \bibinfo {author} {\bibfnamefont {A.}~\bibnamefont {Tiribocchi}},\ }\href@noop {} {\bibfield  {journal} {\bibinfo  {journal} {Comptes Rendus Physique}\ }\textbf {\bibinfo {volume} {16}},\ \bibinfo {pages} {316} (\bibinfo {year} {2015}{\natexlab{a}})}\BibitemShut {NoStop}%
\bibitem [{\citenamefont {Elgeti}\ \emph {et~al.}(2015)\citenamefont {Elgeti}, \citenamefont {Winkler},\ and\ \citenamefont {Gompper}}]{Elgeti15}%
  \BibitemOpen
  \bibfield  {author} {\bibinfo {author} {\bibfnamefont {J.}~\bibnamefont {Elgeti}}, \bibinfo {author} {\bibfnamefont {R.}~\bibnamefont {Winkler}}, \ and\ \bibinfo {author} {\bibfnamefont {G.}~\bibnamefont {Gompper}},\ }\href@noop {} {\bibfield  {journal} {\bibinfo  {journal} {Rep. Prog. Phys.}\ }\textbf {\bibinfo {volume} {78}},\ \bibinfo {pages} {056601} (\bibinfo {year} {2015})}\BibitemShut {NoStop}%
\bibitem [{\citenamefont {Carenza}\ \emph {et~al.}(2019)\citenamefont {Carenza}, \citenamefont {Gonnella}, \citenamefont {Lamura}, \citenamefont {Negro},\ and\ \citenamefont {Tiribocchi}}]{Care2019}%
  \BibitemOpen
  \bibfield  {author} {\bibinfo {author} {\bibfnamefont {L.~N.}\ \bibnamefont {Carenza}}, \bibinfo {author} {\bibfnamefont {G.}~\bibnamefont {Gonnella}}, \bibinfo {author} {\bibfnamefont {A.}~\bibnamefont {Lamura}}, \bibinfo {author} {\bibfnamefont {G.}~\bibnamefont {Negro}}, \ and\ \bibinfo {author} {\bibfnamefont {A.}~\bibnamefont {Tiribocchi}},\ }\href {\doibase 10.1140/epje/i2019-11843-6} {\bibfield  {journal} {\bibinfo  {journal} {The Eur. Phys. J. E}\ }\textbf {\bibinfo {volume} {42}},\ \bibinfo {pages} {81} (\bibinfo {year} {2019})}\BibitemShut {NoStop}%
\bibitem [{\citenamefont {Gompper}\ \emph {et~al.}(2020)\citenamefont {Gompper}, \citenamefont {Winkler}, \citenamefont {Speck}, \citenamefont {Solon}, \citenamefont {Nardini}, \citenamefont {Peruani}, \citenamefont {Löwen}, \citenamefont {Golestanian}, \citenamefont {Kaupp}, \citenamefont {Alvarez}, \citenamefont {Kiørboe}, \citenamefont {Lauga}, \citenamefont {Poon}, \citenamefont {DeSimone}, \citenamefont {Muiños-Landin}, \citenamefont {Fischer}, \citenamefont {Söker}, \citenamefont {Cichos}, \citenamefont {Kapral}, \citenamefont {Gaspard}, \citenamefont {Ripoll}, \citenamefont {Sagues}, \citenamefont {Doostmohammadi}, \citenamefont {Yeomans}, \citenamefont {Aranson}, \citenamefont {Bechinger}, \citenamefont {Stark}, \citenamefont {Hemelrijk}, \citenamefont {Nedelec}, \citenamefont {Sarkar}, \citenamefont {Aryaksama}, \citenamefont {Lacroix}, \citenamefont {Duclos}, \citenamefont {Yashunsky}, \citenamefont {Silberzan}, \citenamefont {Arroyo},\ and\ \citenamefont {Kale}}]{Gompper_2020}%
  \BibitemOpen
  \bibfield  {author} {\bibinfo {author} {\bibfnamefont {G.}~\bibnamefont {Gompper}}, \bibinfo {author} {\bibfnamefont {R.}~\bibnamefont {Winkler}}, \bibinfo {author} {\bibfnamefont {T.}~\bibnamefont {Speck}}, \bibinfo {author} {\bibfnamefont {A.}~\bibnamefont {Solon}}, \bibinfo {author} {\bibfnamefont {C.}~\bibnamefont {Nardini}}, \bibinfo {author} {\bibfnamefont {F.}~\bibnamefont {Peruani}}, \bibinfo {author} {\bibfnamefont {H.}~\bibnamefont {Löwen}}, \bibinfo {author} {\bibfnamefont {R.}~\bibnamefont {Golestanian}}, \bibinfo {author} {\bibfnamefont {U.}~\bibnamefont {Kaupp}}, \bibinfo {author} {\bibfnamefont {L.}~\bibnamefont {Alvarez}}, \bibinfo {author} {\bibfnamefont {T.}~\bibnamefont {Kiørboe}}, \bibinfo {author} {\bibfnamefont {E.}~\bibnamefont {Lauga}}, \bibinfo {author} {\bibfnamefont {W.}~\bibnamefont {Poon}}, \bibinfo {author} {\bibfnamefont {A.}~\bibnamefont {DeSimone}}, \bibinfo {author} {\bibfnamefont {S.}~\bibnamefont {Muiños-Landin}}, \bibinfo {author} {\bibfnamefont {A.}~\bibnamefont
  {Fischer}}, \bibinfo {author} {\bibfnamefont {N.}~\bibnamefont {Söker}}, \bibinfo {author} {\bibfnamefont {F.}~\bibnamefont {Cichos}}, \bibinfo {author} {\bibfnamefont {R.}~\bibnamefont {Kapral}}, \bibinfo {author} {\bibfnamefont {P.}~\bibnamefont {Gaspard}}, \bibinfo {author} {\bibfnamefont {M.}~\bibnamefont {Ripoll}}, \bibinfo {author} {\bibfnamefont {F.}~\bibnamefont {Sagues}}, \bibinfo {author} {\bibfnamefont {A.}~\bibnamefont {Doostmohammadi}}, \bibinfo {author} {\bibfnamefont {J.}~\bibnamefont {Yeomans}}, \bibinfo {author} {\bibfnamefont {I.}~\bibnamefont {Aranson}}, \bibinfo {author} {\bibfnamefont {C.}~\bibnamefont {Bechinger}}, \bibinfo {author} {\bibfnamefont {H.}~\bibnamefont {Stark}}, \bibinfo {author} {\bibfnamefont {C.}~\bibnamefont {Hemelrijk}}, \bibinfo {author} {\bibfnamefont {F.}~\bibnamefont {Nedelec}}, \bibinfo {author} {\bibfnamefont {T.}~\bibnamefont {Sarkar}}, \bibinfo {author} {\bibfnamefont {T.}~\bibnamefont {Aryaksama}}, \bibinfo {author} {\bibfnamefont {M.}~\bibnamefont
  {Lacroix}}, \bibinfo {author} {\bibfnamefont {G.}~\bibnamefont {Duclos}}, \bibinfo {author} {\bibfnamefont {V.}~\bibnamefont {Yashunsky}}, \bibinfo {author} {\bibfnamefont {P.}~\bibnamefont {Silberzan}}, \bibinfo {author} {\bibfnamefont {M.}~\bibnamefont {Arroyo}}, \ and\ \bibinfo {author} {\bibfnamefont {S.}~\bibnamefont {Kale}},\ }\href {\doibase 10.1088/1361-648X/ab6348} {\bibfield  {journal} {\bibinfo  {journal} {J. Phys. Condens. Matter}\ }\textbf {\bibinfo {volume} {32}},\ \bibinfo {pages} {193001} (\bibinfo {year} {2020})}\BibitemShut {NoStop}%
\bibitem [{\citenamefont {Sanchez}\ \emph {et~al.}(2012)\citenamefont {Sanchez}, \citenamefont {Chen}, \citenamefont {DeCamp}, \citenamefont {Heymann},\ and\ \citenamefont {Dogic}}]{sanchez2012}%
  \BibitemOpen
  \bibfield  {author} {\bibinfo {author} {\bibfnamefont {T.}~\bibnamefont {Sanchez}}, \bibinfo {author} {\bibfnamefont {D.}~\bibnamefont {Chen}}, \bibinfo {author} {\bibfnamefont {S.}~\bibnamefont {DeCamp}}, \bibinfo {author} {\bibfnamefont {M.}~\bibnamefont {Heymann}}, \ and\ \bibinfo {author} {\bibfnamefont {Z.}~\bibnamefont {Dogic}},\ }\href@noop {} {\bibfield  {journal} {\bibinfo  {journal} {Nature}\ }\textbf {\bibinfo {volume} {491}},\ \bibinfo {pages} {431} (\bibinfo {year} {2012})}\BibitemShut {NoStop}%
\bibitem [{\citenamefont {{De Magistris}}\ and\ \citenamefont {Marenduzzo}(2015)}]{demagistris2015}%
  \BibitemOpen
  \bibfield  {author} {\bibinfo {author} {\bibfnamefont {G.}~\bibnamefont {{De Magistris}}}\ and\ \bibinfo {author} {\bibfnamefont {D.}~\bibnamefont {Marenduzzo}},\ }\href {\doibase https://doi.org/10.1016/j.physa.2014.06.061} {\bibfield  {journal} {\bibinfo  {journal} {Physica A}\ }\textbf {\bibinfo {volume} {418}},\ \bibinfo {pages} {65} (\bibinfo {year} {2015})},\ \bibinfo {note} {proceedings of the 13th International Summer School on Fundamental Problems in Statistical Physics}\BibitemShut {NoStop}%
\bibitem [{\citenamefont {Cagnetta}\ \emph {et~al.}(2017)\citenamefont {Cagnetta}, \citenamefont {Corberi}, \citenamefont {Gonnella},\ and\ \citenamefont {Suma}}]{cagnetta2017large}%
  \BibitemOpen
  \bibfield  {author} {\bibinfo {author} {\bibfnamefont {F.}~\bibnamefont {Cagnetta}}, \bibinfo {author} {\bibfnamefont {F.}~\bibnamefont {Corberi}}, \bibinfo {author} {\bibfnamefont {G.}~\bibnamefont {Gonnella}}, \ and\ \bibinfo {author} {\bibfnamefont {A.}~\bibnamefont {Suma}},\ }\href@noop {} {\bibfield  {journal} {\bibinfo  {journal} {Phys. Rev. Lett.}\ }\textbf {\bibinfo {volume} {119}},\ \bibinfo {pages} {158002} (\bibinfo {year} {2017})}\BibitemShut {NoStop}%
\bibitem [{\citenamefont {Kumar}\ \emph {et~al.}(2018)\citenamefont {Kumar}, \citenamefont {Zhang}, \citenamefont {de~Pablo},\ and\ \citenamefont {Gardel}}]{kumar2018}%
  \BibitemOpen
  \bibfield  {author} {\bibinfo {author} {\bibfnamefont {N.}~\bibnamefont {Kumar}}, \bibinfo {author} {\bibfnamefont {R.}~\bibnamefont {Zhang}}, \bibinfo {author} {\bibfnamefont {J.}~\bibnamefont {de~Pablo}}, \ and\ \bibinfo {author} {\bibfnamefont {M.}~\bibnamefont {Gardel}},\ }\href@noop {} {\bibfield  {journal} {\bibinfo  {journal} {Sci. Adv.}\ }\textbf {\bibinfo {volume} {4}},\ \bibinfo {pages} {eaat7779} (\bibinfo {year} {2018})}\BibitemShut {NoStop}%
\bibitem [{\citenamefont {Doostmohammadi}\ \emph {et~al.}(2018)\citenamefont {Doostmohammadi}, \citenamefont {Ign{\'e}s-Mullol}, \citenamefont {Yeomans},\ and\ \citenamefont {Sagu{\'e}s}}]{Doostmo2018}%
  \BibitemOpen
  \bibfield  {author} {\bibinfo {author} {\bibfnamefont {A.}~\bibnamefont {Doostmohammadi}}, \bibinfo {author} {\bibfnamefont {J.}~\bibnamefont {Ign{\'e}s-Mullol}}, \bibinfo {author} {\bibfnamefont {J.}~\bibnamefont {Yeomans}}, \ and\ \bibinfo {author} {\bibfnamefont {F.}~\bibnamefont {Sagu{\'e}s}},\ }\href {\doibase 10.1038/s41467-018-05666-8} {\bibfield  {journal} {\bibinfo  {journal} {Nat. Comm.}\ }\textbf {\bibinfo {volume} {9}},\ \bibinfo {pages} {3246} (\bibinfo {year} {2018})}\BibitemShut {NoStop}%
\bibitem [{\citenamefont {Saintillan}(2018)}]{Santillan2018}%
  \BibitemOpen
  \bibfield  {author} {\bibinfo {author} {\bibfnamefont {D.}~\bibnamefont {Saintillan}},\ }\href@noop {} {\bibfield  {journal} {\bibinfo  {journal} {Annu. Rev. Fluid Mech.}\ }\textbf {\bibinfo {volume} {50}},\ \bibinfo {pages} {563} (\bibinfo {year} {2018})}\BibitemShut {NoStop}%
\bibitem [{\citenamefont {Negro}\ \emph {et~al.}(2019)\citenamefont {Negro}, \citenamefont {Lamura}, \citenamefont {Gonnella},\ and\ \citenamefont {Marenduzzo}}]{negro2019hydrodynamics}%
  \BibitemOpen
  \bibfield  {author} {\bibinfo {author} {\bibfnamefont {G.}~\bibnamefont {Negro}}, \bibinfo {author} {\bibfnamefont {A.}~\bibnamefont {Lamura}}, \bibinfo {author} {\bibfnamefont {G.}~\bibnamefont {Gonnella}}, \ and\ \bibinfo {author} {\bibfnamefont {D.}~\bibnamefont {Marenduzzo}},\ }\href@noop {} {\bibfield  {journal} {\bibinfo  {journal} {EPL}\ }\textbf {\bibinfo {volume} {127}},\ \bibinfo {pages} {58001} (\bibinfo {year} {2019})}\BibitemShut {NoStop}%
\bibitem [{\citenamefont {Carenza}\ \emph {et~al.}(2020)\citenamefont {Carenza}, \citenamefont {Gonnella}, \citenamefont {Lamura}, \citenamefont {Marenduzzo}, \citenamefont {Negro},\ and\ \citenamefont {Tiribocchi}}]{carenza2020soft}%
  \BibitemOpen
  \bibfield  {author} {\bibinfo {author} {\bibfnamefont {L.}~\bibnamefont {Carenza}}, \bibinfo {author} {\bibfnamefont {G.}~\bibnamefont {Gonnella}}, \bibinfo {author} {\bibfnamefont {A.}~\bibnamefont {Lamura}}, \bibinfo {author} {\bibfnamefont {D.}~\bibnamefont {Marenduzzo}}, \bibinfo {author} {\bibfnamefont {G.}~\bibnamefont {Negro}}, \ and\ \bibinfo {author} {\bibfnamefont {A.}~\bibnamefont {Tiribocchi}},\ }\href@noop {} {\bibfield  {journal} {\bibinfo  {journal} {Scientific Reports}\ }\textbf {\bibinfo {volume} {10}},\ \bibinfo {pages} {15936} (\bibinfo {year} {2020})}\BibitemShut {NoStop}%
\bibitem [{\citenamefont {Giordano}\ \emph {et~al.}(2021)\citenamefont {Giordano}, \citenamefont {Bonelli}, \citenamefont {Carenza}, \citenamefont {Gonnella},\ and\ \citenamefont {Negro}}]{giordano2021activity}%
  \BibitemOpen
  \bibfield  {author} {\bibinfo {author} {\bibfnamefont {M.~G.}\ \bibnamefont {Giordano}}, \bibinfo {author} {\bibfnamefont {F.}~\bibnamefont {Bonelli}}, \bibinfo {author} {\bibfnamefont {L.~N.}\ \bibnamefont {Carenza}}, \bibinfo {author} {\bibfnamefont {G.}~\bibnamefont {Gonnella}}, \ and\ \bibinfo {author} {\bibfnamefont {G.}~\bibnamefont {Negro}},\ }\href@noop {} {\bibfield  {journal} {\bibinfo  {journal} {Europhysics Letters}\ }\textbf {\bibinfo {volume} {133}},\ \bibinfo {pages} {58004} (\bibinfo {year} {2021})}\BibitemShut {NoStop}%
\bibitem [{\citenamefont {Zhang}\ and\ \citenamefont {Fodor}(2023)}]{Fodor23}%
  \BibitemOpen
  \bibfield  {author} {\bibinfo {author} {\bibfnamefont {Y.}~\bibnamefont {Zhang}}\ and\ \bibinfo {author} {\bibfnamefont {E.}~\bibnamefont {Fodor}},\ }\href@noop {} {\bibfield  {journal} {\bibinfo  {journal} {Phys. Rev. Lett.}\ }\textbf {\bibinfo {volume} {131}},\ \bibinfo {pages} {238302} (\bibinfo {year} {2023})}\BibitemShut {NoStop}%
\bibitem [{\citenamefont {Hernández-López}\ \emph {et~al.}(2023)\citenamefont {Hernández-López}, \citenamefont {Baconnier}, \citenamefont {Coulais}, \citenamefont {Dauchot},\ and\ \citenamefont {D\"uring}}]{Dauchot23}%
  \BibitemOpen
  \bibfield  {author} {\bibinfo {author} {\bibfnamefont {C.}~\bibnamefont {Hernández-López}}, \bibinfo {author} {\bibfnamefont {P.}~\bibnamefont {Baconnier}}, \bibinfo {author} {\bibfnamefont {C.}~\bibnamefont {Coulais}}, \bibinfo {author} {\bibfnamefont {O.}~\bibnamefont {Dauchot}}, \ and\ \bibinfo {author} {\bibfnamefont {G.}~\bibnamefont {D\"uring}},\ }\href@noop {} {\enquote {\bibinfo {title} {Active solids: Rigid body motion and shape-changing mechanisms},}\ } (\bibinfo {year} {2023}),\ \bibinfo {note} {arXiv:2310.12879}\BibitemShut {NoStop}%
\bibitem [{\citenamefont {Caporusso}\ \emph {et~al.}(2023{\natexlab{a}})\citenamefont {Caporusso}, \citenamefont {Gonnella},\ and\ \citenamefont {Levis}}]{Caporusso-chiral23}%
  \BibitemOpen
  \bibfield  {author} {\bibinfo {author} {\bibfnamefont {C.~B.}\ \bibnamefont {Caporusso}}, \bibinfo {author} {\bibfnamefont {G.}~\bibnamefont {Gonnella}}, \ and\ \bibinfo {author} {\bibfnamefont {D.}~\bibnamefont {Levis}},\ }\href@noop {} {\enquote {\bibinfo {title} {Phase coexistence and edge currents in the chiral lennard-jones fluid},}\ } (\bibinfo {year} {2023}{\natexlab{a}}),\ \bibinfo {note} {arXiv:2307.03528}\BibitemShut {NoStop}%
\bibitem [{\citenamefont {Wiese}\ \emph {et~al.}(2023)\citenamefont {Wiese}, \citenamefont {Kroy},\ and\ \citenamefont {Levis}}]{Wiese23}%
  \BibitemOpen
  \bibfield  {author} {\bibinfo {author} {\bibfnamefont {R.}~\bibnamefont {Wiese}}, \bibinfo {author} {\bibfnamefont {K.}~\bibnamefont {Kroy}}, \ and\ \bibinfo {author} {\bibfnamefont {D.}~\bibnamefont {Levis}},\ }\href@noop {} {\bibfield  {journal} {\bibinfo  {journal} {Phys. Rev. Lett.}\ }\textbf {\bibinfo {volume} {131}},\ \bibinfo {pages} {178302} (\bibinfo {year} {2023})}\BibitemShut {NoStop}%
\bibitem [{\citenamefont {Caprini}\ and\ \citenamefont {L\"owen}(2023)}]{Caprini23}%
  \BibitemOpen
  \bibfield  {author} {\bibinfo {author} {\bibfnamefont {L.}~\bibnamefont {Caprini}}\ and\ \bibinfo {author} {\bibfnamefont {H.}~\bibnamefont {L\"owen}},\ }\href@noop {} {\bibfield  {journal} {\bibinfo  {journal} {Phys. Rev. Lett.}\ }\textbf {\bibinfo {volume} {130}},\ \bibinfo {pages} {148202} (\bibinfo {year} {2023})}\BibitemShut {NoStop}%
\bibitem [{\citenamefont {Bayram}\ \emph {et~al.}(2023)\citenamefont {Bayram}, \citenamefont {Schwarzendahl}, \citenamefont {L\"owen},\ and\ \citenamefont {Biancofiore}}]{Bayram23}%
  \BibitemOpen
  \bibfield  {author} {\bibinfo {author} {\bibfnamefont {A.~G.}\ \bibnamefont {Bayram}}, \bibinfo {author} {\bibfnamefont {F.~J.}\ \bibnamefont {Schwarzendahl}}, \bibinfo {author} {\bibfnamefont {H.}~\bibnamefont {L\"owen}}, \ and\ \bibinfo {author} {\bibfnamefont {L.}~\bibnamefont {Biancofiore}},\ }\href@noop {} {\bibfield  {journal} {\bibinfo  {journal} {Soft Matter}\ }\textbf {\bibinfo {volume} {19}},\ \bibinfo {pages} {4571} (\bibinfo {year} {2023})}\BibitemShut {NoStop}%
\bibitem [{\citenamefont {Butt}\ \emph {et~al.}(2010)\citenamefont {Butt}, \citenamefont {Mufti}, \citenamefont {Humayun}, \citenamefont {Rosenthal}, \citenamefont {Khan}, \citenamefont {Khan},\ and\ \citenamefont {Molloy}}]{butt}%
  \BibitemOpen
  \bibfield  {author} {\bibinfo {author} {\bibfnamefont {T.}~\bibnamefont {Butt}}, \bibinfo {author} {\bibfnamefont {T.}~\bibnamefont {Mufti}}, \bibinfo {author} {\bibfnamefont {A.}~\bibnamefont {Humayun}}, \bibinfo {author} {\bibfnamefont {P.~B.}\ \bibnamefont {Rosenthal}}, \bibinfo {author} {\bibfnamefont {S.}~\bibnamefont {Khan}}, \bibinfo {author} {\bibfnamefont {S.}~\bibnamefont {Khan}}, \ and\ \bibinfo {author} {\bibfnamefont {J.~E.}\ \bibnamefont {Molloy}},\ }\href {\doibase 10.1074/jbc.M109.044792} {\bibfield  {journal} {\bibinfo  {journal} {J. Biol. Chem.}\ }\textbf {\bibinfo {volume} {285}},\ \bibinfo {pages} {4964} (\bibinfo {year} {2010})}\BibitemShut {NoStop}%
\bibitem [{\citenamefont {Ballerini}\ \emph {et~al.}(2008)\citenamefont {Ballerini}, \citenamefont {Cabibbo}, \citenamefont {Candelier}, \citenamefont {Cavagna}, \citenamefont {Cisbani}, \citenamefont {Giardina}, \citenamefont {Lecomte}, \citenamefont {Orlandi}, \citenamefont {Parisi}, \citenamefont {Procaccini}, \citenamefont {Viale},\ and\ \citenamefont {Zdravkovic}}]{ballerini}%
  \BibitemOpen
  \bibfield  {author} {\bibinfo {author} {\bibfnamefont {M.}~\bibnamefont {Ballerini}}, \bibinfo {author} {\bibfnamefont {N.}~\bibnamefont {Cabibbo}}, \bibinfo {author} {\bibfnamefont {R.}~\bibnamefont {Candelier}}, \bibinfo {author} {\bibfnamefont {A.}~\bibnamefont {Cavagna}}, \bibinfo {author} {\bibfnamefont {E.}~\bibnamefont {Cisbani}}, \bibinfo {author} {\bibfnamefont {I.}~\bibnamefont {Giardina}}, \bibinfo {author} {\bibfnamefont {V.}~\bibnamefont {Lecomte}}, \bibinfo {author} {\bibfnamefont {A.}~\bibnamefont {Orlandi}}, \bibinfo {author} {\bibfnamefont {G.}~\bibnamefont {Parisi}}, \bibinfo {author} {\bibfnamefont {A.}~\bibnamefont {Procaccini}}, \bibinfo {author} {\bibfnamefont {M.}~\bibnamefont {Viale}}, \ and\ \bibinfo {author} {\bibfnamefont {V.}~\bibnamefont {Zdravkovic}},\ }\href {\doibase 10.1073/pnas.0711437105} {\bibfield  {journal} {\bibinfo  {journal} {Proc. Natl. Acad. Sci. U.S.A}\ }\textbf {\bibinfo {volume} {105}},\ \bibinfo {pages} {1232} (\bibinfo {year} {2008})}\BibitemShut {NoStop}%
\bibitem [{\citenamefont {Palacci}\ \emph {et~al.}(2010)\citenamefont {Palacci}, \citenamefont {Cottin-Bizonne}, \citenamefont {Ybert},\ and\ \citenamefont {Bocquet}}]{Palacci10}%
  \BibitemOpen
  \bibfield  {author} {\bibinfo {author} {\bibfnamefont {J.}~\bibnamefont {Palacci}}, \bibinfo {author} {\bibfnamefont {C.}~\bibnamefont {Cottin-Bizonne}}, \bibinfo {author} {\bibfnamefont {C.}~\bibnamefont {Ybert}}, \ and\ \bibinfo {author} {\bibfnamefont {L.}~\bibnamefont {Bocquet}},\ }\href@noop {} {\bibfield  {journal} {\bibinfo  {journal} {Phys. Rev. Lett.}\ }\textbf {\bibinfo {volume} {105}},\ \bibinfo {pages} {088304} (\bibinfo {year} {2010})}\BibitemShut {NoStop}%
\bibitem [{\citenamefont {Buttinoni}\ \emph {et~al.}(2013)\citenamefont {Buttinoni}, \citenamefont {Bialk\'e}, \citenamefont {K\"ummel}, \citenamefont {L\"owen}, \citenamefont {Bechinger},\ and\ \citenamefont {Speck}}]{Lowen}%
  \BibitemOpen
  \bibfield  {author} {\bibinfo {author} {\bibfnamefont {I.}~\bibnamefont {Buttinoni}}, \bibinfo {author} {\bibfnamefont {J.}~\bibnamefont {Bialk\'e}}, \bibinfo {author} {\bibfnamefont {F.}~\bibnamefont {K\"ummel}}, \bibinfo {author} {\bibfnamefont {H.}~\bibnamefont {L\"owen}}, \bibinfo {author} {\bibfnamefont {C.}~\bibnamefont {Bechinger}}, \ and\ \bibinfo {author} {\bibfnamefont {T.}~\bibnamefont {Speck}},\ }\href {\doibase 10.1103/PhysRevLett.110.238301} {\bibfield  {journal} {\bibinfo  {journal} {Phys. Rev. Lett.}\ }\textbf {\bibinfo {volume} {110}},\ \bibinfo {pages} {238301} (\bibinfo {year} {2013})}\BibitemShut {NoStop}%
\bibitem [{\citenamefont {Narayan}\ \emph {et~al.}(2007)\citenamefont {Narayan}, \citenamefont {Ramaswamy},\ and\ \citenamefont {Menon}}]{Narayan07}%
  \BibitemOpen
  \bibfield  {author} {\bibinfo {author} {\bibfnamefont {V.}~\bibnamefont {Narayan}}, \bibinfo {author} {\bibfnamefont {S.}~\bibnamefont {Ramaswamy}}, \ and\ \bibinfo {author} {\bibfnamefont {N.}~\bibnamefont {Menon}},\ }\href@noop {} {\bibfield  {journal} {\bibinfo  {journal} {Science}\ }\textbf {\bibinfo {volume} {317}},\ \bibinfo {pages} {105} (\bibinfo {year} {2007})}\BibitemShut {NoStop}%
\bibitem [{\citenamefont {Kudrolli}\ \emph {et~al.}(2008)\citenamefont {Kudrolli}, \citenamefont {Lumay}, \citenamefont {Volfson},\ and\ \citenamefont {Tsimring}}]{kudrolli2008swarming}%
  \BibitemOpen
  \bibfield  {author} {\bibinfo {author} {\bibfnamefont {A.}~\bibnamefont {Kudrolli}}, \bibinfo {author} {\bibfnamefont {G.}~\bibnamefont {Lumay}}, \bibinfo {author} {\bibfnamefont {D.}~\bibnamefont {Volfson}}, \ and\ \bibinfo {author} {\bibfnamefont {L.~S.}\ \bibnamefont {Tsimring}},\ }\href@noop {} {\bibfield  {journal} {\bibinfo  {journal} {Phys. Rev. Lett.}\ }\textbf {\bibinfo {volume} {100}},\ \bibinfo {pages} {058001} (\bibinfo {year} {2008})}\BibitemShut {NoStop}%
\bibitem [{\citenamefont {Suma}\ \emph {et~al.}(2014{\natexlab{a}})\citenamefont {Suma}, \citenamefont {Marenduzzo}, \citenamefont {Gonnella},\ and\ \citenamefont {Orlandini}}]{Suma14}%
  \BibitemOpen
  \bibfield  {author} {\bibinfo {author} {\bibfnamefont {A.}~\bibnamefont {Suma}}, \bibinfo {author} {\bibfnamefont {D.}~\bibnamefont {Marenduzzo}}, \bibinfo {author} {\bibfnamefont {G.}~\bibnamefont {Gonnella}}, \ and\ \bibinfo {author} {\bibfnamefont {E.}~\bibnamefont {Orlandini}},\ }\href@noop {} {\bibfield  {journal} {\bibinfo  {journal} {EPL}\ }\textbf {\bibinfo {volume} {108}},\ \bibinfo {pages} {56004} (\bibinfo {year} {2014}{\natexlab{a}})}\BibitemShut {NoStop}%
\bibitem [{\citenamefont {Cugliandolo}\ \emph {et~al.}(2017)\citenamefont {Cugliandolo}, \citenamefont {Digregorio}, \citenamefont {Gonnella},\ and\ \citenamefont {Suma}}]{Cugliandolo2017}%
  \BibitemOpen
  \bibfield  {author} {\bibinfo {author} {\bibfnamefont {L.~F.}\ \bibnamefont {Cugliandolo}}, \bibinfo {author} {\bibfnamefont {P.}~\bibnamefont {Digregorio}}, \bibinfo {author} {\bibfnamefont {G.}~\bibnamefont {Gonnella}}, \ and\ \bibinfo {author} {\bibfnamefont {A.}~\bibnamefont {Suma}},\ }\href@noop {} {\bibfield  {journal} {\bibinfo  {journal} {Phys. Rev. Lett.}\ }\textbf {\bibinfo {volume} {119}},\ \bibinfo {pages} {119} (\bibinfo {year} {2017})}\BibitemShut {NoStop}%
\bibitem [{\citenamefont {Petrelli}\ \emph {et~al.}(2018)\citenamefont {Petrelli}, \citenamefont {Digregorio}, \citenamefont {Cugliandolo}, \citenamefont {Gonnella},\ and\ \citenamefont {Suma}}]{Petrelli18}%
  \BibitemOpen
  \bibfield  {author} {\bibinfo {author} {\bibfnamefont {I.}~\bibnamefont {Petrelli}}, \bibinfo {author} {\bibfnamefont {P.}~\bibnamefont {Digregorio}}, \bibinfo {author} {\bibfnamefont {L.~F.}\ \bibnamefont {Cugliandolo}}, \bibinfo {author} {\bibfnamefont {G.}~\bibnamefont {Gonnella}}, \ and\ \bibinfo {author} {\bibfnamefont {A.}~\bibnamefont {Suma}},\ }\href@noop {} {\bibfield  {journal} {\bibinfo  {journal} {Eur. Phys. J. E}\ }\textbf {\bibinfo {volume} {41}},\ \bibinfo {pages} {128} (\bibinfo {year} {2018})}\BibitemShut {NoStop}%
\bibitem [{\citenamefont {Digregorio}\ \emph {et~al.}(2019)\citenamefont {Digregorio}, \citenamefont {Levis}, \citenamefont {Suma}, \citenamefont {Cugliandolo}, \citenamefont {Gonnella},\ and\ \citenamefont {Pagonabarraga}}]{Digregorio19}%
  \BibitemOpen
  \bibfield  {author} {\bibinfo {author} {\bibfnamefont {P.}~\bibnamefont {Digregorio}}, \bibinfo {author} {\bibfnamefont {D.}~\bibnamefont {Levis}}, \bibinfo {author} {\bibfnamefont {A.}~\bibnamefont {Suma}}, \bibinfo {author} {\bibfnamefont {L.~F.}\ \bibnamefont {Cugliandolo}}, \bibinfo {author} {\bibfnamefont {G.}~\bibnamefont {Gonnella}}, \ and\ \bibinfo {author} {\bibfnamefont {I.}~\bibnamefont {Pagonabarraga}},\ }\href@noop {} {\bibfield  {journal} {\bibinfo  {journal} {J. Phys. Conf. Series}\ }\textbf {\bibinfo {volume} {1163}},\ \bibinfo {pages} {012073} (\bibinfo {year} {2019})}\BibitemShut {NoStop}%
\bibitem [{\citenamefont {Fily}\ and\ \citenamefont {Marchetti}(2012)}]{Fily12}%
  \BibitemOpen
  \bibfield  {author} {\bibinfo {author} {\bibfnamefont {Y.}~\bibnamefont {Fily}}\ and\ \bibinfo {author} {\bibfnamefont {M.~C.}\ \bibnamefont {Marchetti}},\ }\href@noop {} {\bibfield  {journal} {\bibinfo  {journal} {Phys. Rev. Lett.}\ }\textbf {\bibinfo {volume} {108}},\ \bibinfo {pages} {235702} (\bibinfo {year} {2012})}\BibitemShut {NoStop}%
\bibitem [{\citenamefont {Redner}\ \emph {et~al.}(2013)\citenamefont {Redner}, \citenamefont {Hagan},\ and\ \citenamefont {Baskaran}}]{Redner13}%
  \BibitemOpen
  \bibfield  {author} {\bibinfo {author} {\bibfnamefont {G.~S.}\ \bibnamefont {Redner}}, \bibinfo {author} {\bibfnamefont {M.~F.}\ \bibnamefont {Hagan}}, \ and\ \bibinfo {author} {\bibfnamefont {A.}~\bibnamefont {Baskaran}},\ }\href@noop {} {\bibfield  {journal} {\bibinfo  {journal} {Phys. Rev. Lett.}\ }\textbf {\bibinfo {volume} {110}},\ \bibinfo {pages} {055701} (\bibinfo {year} {2013})}\BibitemShut {NoStop}%
\bibitem [{\citenamefont {Negro}\ \emph {et~al.}(2022)\citenamefont {Negro}, \citenamefont {Caporusso}, \citenamefont {Digregorio}, \citenamefont {Gonnella}, \citenamefont {Lamura},\ and\ \citenamefont {Suma}}]{negro2022hydrodynamic}%
  \BibitemOpen
  \bibfield  {author} {\bibinfo {author} {\bibfnamefont {G.}~\bibnamefont {Negro}}, \bibinfo {author} {\bibfnamefont {C.~B.}\ \bibnamefont {Caporusso}}, \bibinfo {author} {\bibfnamefont {P.}~\bibnamefont {Digregorio}}, \bibinfo {author} {\bibfnamefont {G.}~\bibnamefont {Gonnella}}, \bibinfo {author} {\bibfnamefont {A.}~\bibnamefont {Lamura}}, \ and\ \bibinfo {author} {\bibfnamefont {A.}~\bibnamefont {Suma}},\ }\href@noop {} {\bibfield  {journal} {\bibinfo  {journal} {Eur. Phys. J. E}\ }\textbf {\bibinfo {volume} {45}},\ \bibinfo {pages} {75} (\bibinfo {year} {2022})}\BibitemShut {NoStop}%
\bibitem [{\citenamefont {Caporusso}\ \emph {et~al.}(2020)\citenamefont {Caporusso}, \citenamefont {Digregorio}, \citenamefont {Levis}, \citenamefont {Cugliandolo},\ and\ \citenamefont {Gonnella}}]{Caporusso20}%
  \BibitemOpen
  \bibfield  {author} {\bibinfo {author} {\bibfnamefont {C.~B.}\ \bibnamefont {Caporusso}}, \bibinfo {author} {\bibfnamefont {L.}~\bibnamefont {Digregorio}}, \bibinfo {author} {\bibfnamefont {D.}~\bibnamefont {Levis}}, \bibinfo {author} {\bibfnamefont {L.~F.}\ \bibnamefont {Cugliandolo}}, \ and\ \bibinfo {author} {\bibfnamefont {G.}~\bibnamefont {Gonnella}},\ }\href@noop {} {\bibfield  {journal} {\bibinfo  {journal} {Phys. Rev. Lett.}\ }\textbf {\bibinfo {volume} {125}},\ \bibinfo {pages} {178004} (\bibinfo {year} {2020})}\BibitemShut {NoStop}%
\bibitem [{\citenamefont {Caporusso}\ \emph {et~al.}(2023{\natexlab{b}})\citenamefont {Caporusso}, \citenamefont {Cugliandolo}, \citenamefont {Digregorio}, \citenamefont {Gonnella}, \citenamefont {Levis},\ and\ \citenamefont {Suma}}]{Caporusso22}%
  \BibitemOpen
  \bibfield  {author} {\bibinfo {author} {\bibfnamefont {C.}~\bibnamefont {Caporusso}}, \bibinfo {author} {\bibfnamefont {L.~F.}\ \bibnamefont {Cugliandolo}}, \bibinfo {author} {\bibfnamefont {P.}~\bibnamefont {Digregorio}}, \bibinfo {author} {\bibfnamefont {G.}~\bibnamefont {Gonnella}}, \bibinfo {author} {\bibfnamefont {D.}~\bibnamefont {Levis}}, \ and\ \bibinfo {author} {\bibfnamefont {A.}~\bibnamefont {Suma}},\ }\href@noop {} {\bibfield  {journal} {\bibinfo  {journal} {Phys. Rev. Lett.}\ }\textbf {\bibinfo {volume} {131}},\ \bibinfo {pages} {068201} (\bibinfo {year} {2023}{\natexlab{b}})}\BibitemShut {NoStop}%
\bibitem [{\citenamefont {Moran}\ \emph {et~al.}(2022)\citenamefont {Moran}, \citenamefont {Bruss}, \citenamefont {Sch\"onh\"ofer},\ and\ \citenamefont {Glotzer}}]{Moran22}%
  \BibitemOpen
  \bibfield  {author} {\bibinfo {author} {\bibfnamefont {S.~E.}\ \bibnamefont {Moran}}, \bibinfo {author} {\bibfnamefont {I.~R.}\ \bibnamefont {Bruss}}, \bibinfo {author} {\bibfnamefont {P.~W.~A.}\ \bibnamefont {Sch\"onh\"ofer}}, \ and\ \bibinfo {author} {\bibfnamefont {S.~C.}\ \bibnamefont {Glotzer}},\ }\href@noop {} {\bibfield  {journal} {\bibinfo  {journal} {Soft Matter}\ }\textbf {\bibinfo {volume} {18}},\ \bibinfo {pages} {1044} (\bibinfo {year} {2022})}\BibitemShut {NoStop}%
\bibitem [{\citenamefont {Solon}\ \emph {et~al.}(2015)\citenamefont {Solon}, \citenamefont {Fily}, \citenamefont {Baskaran}, \citenamefont {Cates}, \citenamefont {Kafri}, \citenamefont {Kardar},\ and\ \citenamefont {Tailleur}}]{solon2015pressure}%
  \BibitemOpen
  \bibfield  {author} {\bibinfo {author} {\bibfnamefont {A.~P.}\ \bibnamefont {Solon}}, \bibinfo {author} {\bibfnamefont {Y.}~\bibnamefont {Fily}}, \bibinfo {author} {\bibfnamefont {A.}~\bibnamefont {Baskaran}}, \bibinfo {author} {\bibfnamefont {M.~E.}\ \bibnamefont {Cates}}, \bibinfo {author} {\bibfnamefont {Y.}~\bibnamefont {Kafri}}, \bibinfo {author} {\bibfnamefont {M.}~\bibnamefont {Kardar}}, \ and\ \bibinfo {author} {\bibfnamefont {J.}~\bibnamefont {Tailleur}},\ }\href@noop {} {\bibfield  {journal} {\bibinfo  {journal} {Nature Phys.}\ }\textbf {\bibinfo {volume} {11}},\ \bibinfo {pages} {673} (\bibinfo {year} {2015})}\BibitemShut {NoStop}%
\bibitem [{\citenamefont {Fily}\ \emph {et~al.}(2018)\citenamefont {Fily}, \citenamefont {Kafri}, \citenamefont {Solon}, \citenamefont {Tailleur},\ and\ \citenamefont {Turner}}]{Fily18}%
  \BibitemOpen
  \bibfield  {author} {\bibinfo {author} {\bibfnamefont {Y.}~\bibnamefont {Fily}}, \bibinfo {author} {\bibfnamefont {Y.}~\bibnamefont {Kafri}}, \bibinfo {author} {\bibfnamefont {A.~P.}\ \bibnamefont {Solon}}, \bibinfo {author} {\bibfnamefont {J.}~\bibnamefont {Tailleur}}, \ and\ \bibinfo {author} {\bibfnamefont {A.}~\bibnamefont {Turner}},\ }\href@noop {} {\bibfield  {journal} {\bibinfo  {journal} {J. Phys. A: Math. Theor.}\ }\textbf {\bibinfo {volume} {51}},\ \bibinfo {pages} {044003} (\bibinfo {year} {2018})}\BibitemShut {NoStop}%
\bibitem [{\citenamefont {Pirhadi}\ \emph {et~al.}(2021)\citenamefont {Pirhadi}, \citenamefont {Cheng},\ and\ \citenamefont {Yong}}]{Pirhadi21}%
  \BibitemOpen
  \bibfield  {author} {\bibinfo {author} {\bibfnamefont {E.}~\bibnamefont {Pirhadi}}, \bibinfo {author} {\bibfnamefont {X.}~\bibnamefont {Cheng}}, \ and\ \bibinfo {author} {\bibfnamefont {X.}~\bibnamefont {Yong}},\ }\href@noop {} {\bibfield  {journal} {\bibinfo  {journal} {Scientific Reports}\ }\textbf {\bibinfo {volume} {11}},\ \bibinfo {pages} {22204} (\bibinfo {year} {2021})}\BibitemShut {NoStop}%
\bibitem [{\citenamefont {Joyeux}\ and\ \citenamefont {Bertin}(2016)}]{Joyeux16}%
  \BibitemOpen
  \bibfield  {author} {\bibinfo {author} {\bibfnamefont {M.}~\bibnamefont {Joyeux}}\ and\ \bibinfo {author} {\bibfnamefont {E.}~\bibnamefont {Bertin}},\ }\href@noop {} {\bibfield  {journal} {\bibinfo  {journal} {Phys. Rev. E}\ }\textbf {\bibinfo {volume} {93}},\ \bibinfo {pages} {032605} (\bibinfo {year} {2016})}\BibitemShut {NoStop}%
\bibitem [{\citenamefont {Joyeux}(2017)}]{Joyeux17}%
  \BibitemOpen
  \bibfield  {author} {\bibinfo {author} {\bibfnamefont {M.}~\bibnamefont {Joyeux}},\ }\href@noop {} {\bibfield  {journal} {\bibinfo  {journal} {Phys. Rev. E}\ }\textbf {\bibinfo {volume} {95}},\ \bibinfo {pages} {052603} (\bibinfo {year} {2017})}\BibitemShut {NoStop}%
\bibitem [{\citenamefont {Caporusso}\ \emph {et~al.}(2024)\citenamefont {Caporusso}, \citenamefont {Negro}, \citenamefont {Suma}, \citenamefont {Digregorio}, \citenamefont {Carenza}, \citenamefont {Gonnella},\ and\ \citenamefont {Cugliandolo}}]{Caporusso23}%
  \BibitemOpen
  \bibfield  {author} {\bibinfo {author} {\bibfnamefont {C.~B.}\ \bibnamefont {Caporusso}}, \bibinfo {author} {\bibfnamefont {G.}~\bibnamefont {Negro}}, \bibinfo {author} {\bibfnamefont {A.}~\bibnamefont {Suma}}, \bibinfo {author} {\bibfnamefont {P.}~\bibnamefont {Digregorio}}, \bibinfo {author} {\bibfnamefont {L.~N.}\ \bibnamefont {Carenza}}, \bibinfo {author} {\bibfnamefont {G.}~\bibnamefont {Gonnella}}, \ and\ \bibinfo {author} {\bibfnamefont {L.~F.}\ \bibnamefont {Cugliandolo}},\ }\href {\doibase 10.1039/D3SM01030A} {\bibfield  {journal} {\bibinfo  {journal} {Soft Matter}\ }\textbf {\bibinfo {volume} {20}},\ \bibinfo {pages} {923} (\bibinfo {year} {2024})}\BibitemShut {NoStop}%
\bibitem [{\citenamefont {L\"owen}(2018)}]{Lowen18}%
  \BibitemOpen
  \bibfield  {author} {\bibinfo {author} {\bibfnamefont {H.}~\bibnamefont {L\"owen}},\ }\href@noop {} {\bibfield  {journal} {\bibinfo  {journal} {EPL}\ }\textbf {\bibinfo {volume} {121}},\ \bibinfo {pages} {58001} (\bibinfo {year} {2018})}\BibitemShut {NoStop}%
\bibitem [{\citenamefont {Mallory}\ \emph {et~al.}(2018)\citenamefont {Mallory}, \citenamefont {Valeriani},\ and\ \citenamefont {Cacciuto}}]{Mallory18}%
  \BibitemOpen
  \bibfield  {author} {\bibinfo {author} {\bibfnamefont {S.~A.}\ \bibnamefont {Mallory}}, \bibinfo {author} {\bibfnamefont {C.}~\bibnamefont {Valeriani}}, \ and\ \bibinfo {author} {\bibfnamefont {A.}~\bibnamefont {Cacciuto}},\ }\href@noop {} {\bibfield  {journal} {\bibinfo  {journal} {Annu. Rev. Phys. Chem.}\ }\textbf {\bibinfo {volume} {69}},\ \bibinfo {pages} {59} (\bibinfo {year} {2018})}\BibitemShut {NoStop}%
\bibitem [{\citenamefont {Schwarzendahl}\ \emph {et~al.}(2023)\citenamefont {Schwarzendahl}, \citenamefont {Maulean-Amieva}, \citenamefont {Royall},\ and\ \citenamefont {L\"owen}}]{Paddy23}%
  \BibitemOpen
  \bibfield  {author} {\bibinfo {author} {\bibfnamefont {F.~J.}\ \bibnamefont {Schwarzendahl}}, \bibinfo {author} {\bibfnamefont {A.}~\bibnamefont {Maulean-Amieva}}, \bibinfo {author} {\bibfnamefont {C.~P.}\ \bibnamefont {Royall}}, \ and\ \bibinfo {author} {\bibfnamefont {H.}~\bibnamefont {L\"owen}},\ }\href@noop {} {\bibfield  {journal} {\bibinfo  {journal} {Phys. Rev. E}\ }\textbf {\bibinfo {volume} {107}},\ \bibinfo {pages} {054606} (\bibinfo {year} {2023})}\BibitemShut {NoStop}%
\bibitem [{\citenamefont {Cates}\ and\ \citenamefont {Tailleur}(2015)}]{Cates15}%
  \BibitemOpen
  \bibfield  {author} {\bibinfo {author} {\bibfnamefont {M.~E.}\ \bibnamefont {Cates}}\ and\ \bibinfo {author} {\bibfnamefont {J.}~\bibnamefont {Tailleur}},\ }\href@noop {} {\bibfield  {journal} {\bibinfo  {journal} {Annu. Rev. Cond. Matt. Phys.}\ }\textbf {\bibinfo {volume} {6}},\ \bibinfo {pages} {219} (\bibinfo {year} {2015})}\BibitemShut {NoStop}%
\bibitem [{\citenamefont {Gonnella}\ \emph {et~al.}(2015{\natexlab{b}})\citenamefont {Gonnella}, \citenamefont {Marenduzzo}, \citenamefont {Suma},\ and\ \citenamefont {Tiribocchi}}]{Gonnella15}%
  \BibitemOpen
  \bibfield  {author} {\bibinfo {author} {\bibfnamefont {G.}~\bibnamefont {Gonnella}}, \bibinfo {author} {\bibfnamefont {D.}~\bibnamefont {Marenduzzo}}, \bibinfo {author} {\bibfnamefont {A.}~\bibnamefont {Suma}}, \ and\ \bibinfo {author} {\bibfnamefont {A.}~\bibnamefont {Tiribocchi}},\ }\href@noop {} {\bibfield  {journal} {\bibinfo  {journal} {Comptes Rendus Acad. Sc.}\ }\textbf {\bibinfo {volume} {16}},\ \bibinfo {pages} {316} (\bibinfo {year} {2015}{\natexlab{b}})}\BibitemShut {NoStop}%
\bibitem [{\citenamefont {Digregorio}\ \emph {et~al.}(2018)\citenamefont {Digregorio}, \citenamefont {Levis}, \citenamefont {Suma}, \citenamefont {Cugliandolo}, \citenamefont {Gonnella},\ and\ \citenamefont {Pagonabarraga}}]{Digregorio18}%
  \BibitemOpen
  \bibfield  {author} {\bibinfo {author} {\bibfnamefont {P.}~\bibnamefont {Digregorio}}, \bibinfo {author} {\bibfnamefont {D.}~\bibnamefont {Levis}}, \bibinfo {author} {\bibfnamefont {A.}~\bibnamefont {Suma}}, \bibinfo {author} {\bibfnamefont {L.~F.}\ \bibnamefont {Cugliandolo}}, \bibinfo {author} {\bibfnamefont {G.}~\bibnamefont {Gonnella}}, \ and\ \bibinfo {author} {\bibfnamefont {I.}~\bibnamefont {Pagonabarraga}},\ }\href@noop {} {\bibfield  {journal} {\bibinfo  {journal} {Phys. Rev. Lett.}\ }\textbf {\bibinfo {volume} {121}},\ \bibinfo {pages} {098003} (\bibinfo {year} {2018})}\BibitemShut {NoStop}%
\bibitem [{\citenamefont {Peruani}\ \emph {et~al.}(2006)\citenamefont {Peruani}, \citenamefont {Deutsch},\ and\ \citenamefont {B\"ar}}]{Peruani2006}%
  \BibitemOpen
  \bibfield  {author} {\bibinfo {author} {\bibfnamefont {F.}~\bibnamefont {Peruani}}, \bibinfo {author} {\bibfnamefont {A.}~\bibnamefont {Deutsch}}, \ and\ \bibinfo {author} {\bibfnamefont {M.}~\bibnamefont {B\"ar}},\ }\href@noop {} {\bibfield  {journal} {\bibinfo  {journal} {Phys. Rev. E}\ }\textbf {\bibinfo {volume} {74}},\ \bibinfo {pages} {030904(R)} (\bibinfo {year} {2006})}\BibitemShut {NoStop}%
\bibitem [{\citenamefont {Ginelli}\ \emph {et~al.}(2010)\citenamefont {Ginelli}, \citenamefont {Peruani}, \citenamefont {B\"ar},\ and\ \citenamefont {Chat\'e}}]{Ginelli2010}%
  \BibitemOpen
  \bibfield  {author} {\bibinfo {author} {\bibfnamefont {F.}~\bibnamefont {Ginelli}}, \bibinfo {author} {\bibfnamefont {F.}~\bibnamefont {Peruani}}, \bibinfo {author} {\bibfnamefont {M.}~\bibnamefont {B\"ar}}, \ and\ \bibinfo {author} {\bibfnamefont {H.}~\bibnamefont {Chat\'e}},\ }\href@noop {} {\bibfield  {journal} {\bibinfo  {journal} {Phys. Rev. Lett.}\ }\textbf {\bibinfo {volume} {104}},\ \bibinfo {pages} {184502} (\bibinfo {year} {2010})}\BibitemShut {NoStop}%
\bibitem [{\citenamefont {Yang}\ \emph {et~al.}(2010)\citenamefont {Yang}, \citenamefont {Marceau},\ and\ \citenamefont {Gompper}}]{Yang2010}%
  \BibitemOpen
  \bibfield  {author} {\bibinfo {author} {\bibfnamefont {Y.}~\bibnamefont {Yang}}, \bibinfo {author} {\bibfnamefont {V.}~\bibnamefont {Marceau}}, \ and\ \bibinfo {author} {\bibfnamefont {G.}~\bibnamefont {Gompper}},\ }\href@noop {} {\bibfield  {journal} {\bibinfo  {journal} {Phys. Rev. E}\ }\textbf {\bibinfo {volume} {82}},\ \bibinfo {pages} {031904} (\bibinfo {year} {2010})}\BibitemShut {NoStop}%
\bibitem [{\citenamefont {van Damme}\ \emph {et~al.}(2019)\citenamefont {van Damme}, \citenamefont {Rodenburg}, \citenamefont {van Roij},\ and\ \citenamefont {Dijkstra}}]{Dijkstra19}%
  \BibitemOpen
  \bibfield  {author} {\bibinfo {author} {\bibfnamefont {R.}~\bibnamefont {van Damme}}, \bibinfo {author} {\bibfnamefont {J.}~\bibnamefont {Rodenburg}}, \bibinfo {author} {\bibfnamefont {R.}~\bibnamefont {van Roij}}, \ and\ \bibinfo {author} {\bibfnamefont {M.}~\bibnamefont {Dijkstra}},\ }\href@noop {} {\bibfield  {journal} {\bibinfo  {journal} {J. Chem. Phys.}\ }\textbf {\bibinfo {volume} {150}},\ \bibinfo {pages} {164501} (\bibinfo {year} {2019})}\BibitemShut {NoStop}%
\bibitem [{\citenamefont {B\"ar}\ \emph {et~al.}(2020)\citenamefont {B\"ar}, \citenamefont {Grossmann}, \citenamefont {Heidenreich},\ and\ \citenamefont {Peruani}}]{Bar20}%
  \BibitemOpen
  \bibfield  {author} {\bibinfo {author} {\bibfnamefont {M.}~\bibnamefont {B\"ar}}, \bibinfo {author} {\bibfnamefont {R.}~\bibnamefont {Grossmann}}, \bibinfo {author} {\bibfnamefont {S.}~\bibnamefont {Heidenreich}}, \ and\ \bibinfo {author} {\bibfnamefont {F.}~\bibnamefont {Peruani}},\ }\href@noop {} {\bibfield  {journal} {\bibinfo  {journal} {Annu. Rev. Cond. Matt. Phys.}\ }\textbf {\bibinfo {volume} {11}},\ \bibinfo {pages} {441} (\bibinfo {year} {2020})}\BibitemShut {NoStop}%
\bibitem [{\citenamefont {Suma}\ \emph {et~al.}(2014{\natexlab{b}})\citenamefont {Suma}, \citenamefont {Gonnella}, \citenamefont {Laghezza}, \citenamefont {Lamura}, \citenamefont {Mossa},\ and\ \citenamefont {Cugliandolo}}]{Suma14b}%
  \BibitemOpen
  \bibfield  {author} {\bibinfo {author} {\bibfnamefont {A.}~\bibnamefont {Suma}}, \bibinfo {author} {\bibfnamefont {G.}~\bibnamefont {Gonnella}}, \bibinfo {author} {\bibfnamefont {G.}~\bibnamefont {Laghezza}}, \bibinfo {author} {\bibfnamefont {A.}~\bibnamefont {Lamura}}, \bibinfo {author} {\bibfnamefont {A.}~\bibnamefont {Mossa}}, \ and\ \bibinfo {author} {\bibfnamefont {L.~F.}\ \bibnamefont {Cugliandolo}},\ }\href@noop {} {\bibfield  {journal} {\bibinfo  {journal} {Phys. Rev. E}\ }\textbf {\bibinfo {volume} {90}},\ \bibinfo {pages} {052130} (\bibinfo {year} {2014}{\natexlab{b}})}\BibitemShut {NoStop}%
\bibitem [{\citenamefont {Cugliandolo}\ \emph {et~al.}(2015)\citenamefont {Cugliandolo}, \citenamefont {Gonnella},\ and\ \citenamefont {Suma}}]{Suma14c}%
  \BibitemOpen
  \bibfield  {author} {\bibinfo {author} {\bibfnamefont {L.~F.}\ \bibnamefont {Cugliandolo}}, \bibinfo {author} {\bibfnamefont {G.}~\bibnamefont {Gonnella}}, \ and\ \bibinfo {author} {\bibfnamefont {A.}~\bibnamefont {Suma}},\ }\href@noop {} {\bibfield  {journal} {\bibinfo  {journal} {Phys. Rev. E}\ }\textbf {\bibinfo {volume} {91}},\ \bibinfo {pages} {062124} (\bibinfo {year} {2015})}\BibitemShut {NoStop}%
\bibitem [{\citenamefont {Suma}\ \emph {et~al.}(2016)\citenamefont {Suma}, \citenamefont {Cugliandolo},\ and\ \citenamefont {Gonnella}}]{Suma16}%
  \BibitemOpen
  \bibfield  {author} {\bibinfo {author} {\bibfnamefont {A.}~\bibnamefont {Suma}}, \bibinfo {author} {\bibfnamefont {L.~F.}\ \bibnamefont {Cugliandolo}}, \ and\ \bibinfo {author} {\bibfnamefont {G.}~\bibnamefont {Gonnella}},\ }\href {http://stacks.iop.org/1742-5468/2016/i=5/a=054029} {\bibfield  {journal} {\bibinfo  {journal} {J. Stat. Mech.}\ }\textbf {\bibinfo {volume} {2016}},\ \bibinfo {pages} {054029} (\bibinfo {year} {2016})}\BibitemShut {NoStop}%
\bibitem [{\citenamefont {Schwarz-Linek}\ \emph {et~al.}(2012)\citenamefont {Schwarz-Linek}, \citenamefont {Valeriani}, \citenamefont {Cacciuto}, \citenamefont {Cates}, \citenamefont {Marenduzzo}, \citenamefont {Morozov},\ and\ \citenamefont {Poon}}]{Schwarz-Linek12}%
  \BibitemOpen
  \bibfield  {author} {\bibinfo {author} {\bibfnamefont {J.}~\bibnamefont {Schwarz-Linek}}, \bibinfo {author} {\bibfnamefont {C.}~\bibnamefont {Valeriani}}, \bibinfo {author} {\bibfnamefont {A.}~\bibnamefont {Cacciuto}}, \bibinfo {author} {\bibfnamefont {M.}~\bibnamefont {Cates}}, \bibinfo {author} {\bibfnamefont {D.}~\bibnamefont {Marenduzzo}}, \bibinfo {author} {\bibfnamefont {A.}~\bibnamefont {Morozov}}, \ and\ \bibinfo {author} {\bibfnamefont {W.}~\bibnamefont {Poon}},\ }\href@noop {} {\bibfield  {journal} {\bibinfo  {journal} {Proc. Natl. Acad. Sci. U. S. A.}\ }\textbf {\bibinfo {volume} {109}},\ \bibinfo {pages} {4052} (\bibinfo {year} {2012})}\BibitemShut {NoStop}%
\bibitem [{\citenamefont {Tung}\ \emph {et~al.}(2016)\citenamefont {Tung}, \citenamefont {Harder}, \citenamefont {Valeriani},\ and\ \citenamefont {Cacciuto}}]{Tung16}%
  \BibitemOpen
  \bibfield  {author} {\bibinfo {author} {\bibfnamefont {C.}~\bibnamefont {Tung}}, \bibinfo {author} {\bibfnamefont {J.}~\bibnamefont {Harder}}, \bibinfo {author} {\bibfnamefont {C.}~\bibnamefont {Valeriani}}, \ and\ \bibinfo {author} {\bibfnamefont {A.}~\bibnamefont {Cacciuto}},\ }\href@noop {} {\bibfield  {journal} {\bibinfo  {journal} {Soft Matter}\ }\textbf {\bibinfo {volume} {12}},\ \bibinfo {pages} {555} (\bibinfo {year} {2016})}\BibitemShut {NoStop}%
\bibitem [{\citenamefont {Petrelli}\ \emph {et~al.}(2020)\citenamefont {Petrelli}, \citenamefont {Cugliandolo}, \citenamefont {Gonnella},\ and\ \citenamefont {Suma}}]{Petrelli20}%
  \BibitemOpen
  \bibfield  {author} {\bibinfo {author} {\bibfnamefont {I.}~\bibnamefont {Petrelli}}, \bibinfo {author} {\bibfnamefont {L.~F.}\ \bibnamefont {Cugliandolo}}, \bibinfo {author} {\bibfnamefont {G.}~\bibnamefont {Gonnella}}, \ and\ \bibinfo {author} {\bibfnamefont {A.}~\bibnamefont {Suma}},\ }\href@noop {} {\bibfield  {journal} {\bibinfo  {journal} {Phys. Rev. E}\ }\textbf {\bibinfo {volume} {102}},\ \bibinfo {pages} {012609} (\bibinfo {year} {2020})}\BibitemShut {NoStop}%
\bibitem [{\citenamefont {Winkler}(2016)}]{Winkler2016a}%
  \BibitemOpen
  \bibfield  {author} {\bibinfo {author} {\bibfnamefont {R.~G.}\ \bibnamefont {Winkler}},\ }\href {\doibase 10.1039/C5SM02965A} {\bibfield  {journal} {\bibinfo  {journal} {Soft Matter}\ }\textbf {\bibinfo {volume} {12}},\ \bibinfo {pages} {3737} (\bibinfo {year} {2016})}\BibitemShut {NoStop}%
\bibitem [{\citenamefont {Clop\'es}\ \emph {et~al.}(2020)\citenamefont {Clop\'es}, \citenamefont {Gompper},\ and\ \citenamefont {Winkler}}]{Coples}%
  \BibitemOpen
  \bibfield  {author} {\bibinfo {author} {\bibfnamefont {J.}~\bibnamefont {Clop\'es}}, \bibinfo {author} {\bibfnamefont {G.}~\bibnamefont {Gompper}}, \ and\ \bibinfo {author} {\bibfnamefont {R.~G.}\ \bibnamefont {Winkler}},\ }\href@noop {} {\bibfield  {journal} {\bibinfo  {journal} {Soft Matter}\ }\textbf {\bibinfo {volume} {16}},\ \bibinfo {pages} {10676} (\bibinfo {year} {2020})}\BibitemShut {NoStop}%
\bibitem [{\citenamefont {Hargus}\ \emph {et~al.}(2020)\citenamefont {Hargus}, \citenamefont {Klymko}, \citenamefont {Epstein},\ and\ \citenamefont {Mandadapu}}]{Hargus20}%
  \BibitemOpen
  \bibfield  {author} {\bibinfo {author} {\bibfnamefont {C.}~\bibnamefont {Hargus}}, \bibinfo {author} {\bibfnamefont {K.}~\bibnamefont {Klymko}}, \bibinfo {author} {\bibfnamefont {J.~M.}\ \bibnamefont {Epstein}}, \ and\ \bibinfo {author} {\bibfnamefont {K.~K.}\ \bibnamefont {Mandadapu}},\ }\href@noop {} {\bibfield  {journal} {\bibinfo  {journal} {J. Chem. Phys.}\ }\textbf {\bibinfo {volume} {152}},\ \bibinfo {pages} {201102} (\bibinfo {year} {2020})}\BibitemShut {NoStop}%
\bibitem [{\citenamefont {Mandal}\ \emph {et~al.}(2017)\citenamefont {Mandal}, \citenamefont {Bhuyan}, \citenamefont {Chaudhuri}, \citenamefont {Rao},\ and\ \citenamefont {Dasgupta}}]{Mandal17}%
  \BibitemOpen
  \bibfield  {author} {\bibinfo {author} {\bibfnamefont {R.}~\bibnamefont {Mandal}}, \bibinfo {author} {\bibfnamefont {P.~J.}\ \bibnamefont {Bhuyan}}, \bibinfo {author} {\bibfnamefont {P.}~\bibnamefont {Chaudhuri}}, \bibinfo {author} {\bibfnamefont {M.}~\bibnamefont {Rao}}, \ and\ \bibinfo {author} {\bibfnamefont {C.}~\bibnamefont {Dasgupta}},\ }\href@noop {} {\bibfield  {journal} {\bibinfo  {journal} {Phys. Rev. E}\ }\textbf {\bibinfo {volume} {96}},\ \bibinfo {pages} {042605} (\bibinfo {year} {2017})}\BibitemShut {NoStop}%
\bibitem [{\citenamefont {Nandi}\ \emph {et~al.}(2018)\citenamefont {Nandi}, \citenamefont {Mandal}, \citenamefont {Bhuyan}, \citenamefont {Dasgupta}, \citenamefont {Rao},\ and\ \citenamefont {Gov}}]{Mandal2018}%
  \BibitemOpen
  \bibfield  {author} {\bibinfo {author} {\bibfnamefont {S.~K.}\ \bibnamefont {Nandi}}, \bibinfo {author} {\bibfnamefont {R.}~\bibnamefont {Mandal}}, \bibinfo {author} {\bibfnamefont {P.~J.}\ \bibnamefont {Bhuyan}}, \bibinfo {author} {\bibfnamefont {C.}~\bibnamefont {Dasgupta}}, \bibinfo {author} {\bibfnamefont {M.}~\bibnamefont {Rao}}, \ and\ \bibinfo {author} {\bibfnamefont {N.~S.}\ \bibnamefont {Gov}},\ }\href@noop {} {\bibfield  {journal} {\bibinfo  {journal} {Proc. Natl. Acad. Sci. U.S.A}\ }\textbf {\bibinfo {volume} {115}},\ \bibinfo {pages} {7688} (\bibinfo {year} {2018})}\BibitemShut {NoStop}%
\bibitem [{\citenamefont {Siebert}\ \emph {et~al.}(2017)\citenamefont {Siebert}, \citenamefont {Letz}, \citenamefont {Speck},\ and\ \citenamefont {Virnau}}]{Siebert2017}%
  \BibitemOpen
  \bibfield  {author} {\bibinfo {author} {\bibfnamefont {J.~T.}\ \bibnamefont {Siebert}}, \bibinfo {author} {\bibfnamefont {J.}~\bibnamefont {Letz}}, \bibinfo {author} {\bibfnamefont {T.}~\bibnamefont {Speck}}, \ and\ \bibinfo {author} {\bibfnamefont {P.}~\bibnamefont {Virnau}},\ }\href@noop {} {\bibfield  {journal} {\bibinfo  {journal} {Soft Matter}\ }\textbf {\bibinfo {volume} {13}},\ \bibinfo {pages} {1020} (\bibinfo {year} {2017})}\BibitemShut {NoStop}%
\bibitem [{\citenamefont {Venkatareddy}\ \emph {et~al.}(2023)\citenamefont {Venkatareddy}, \citenamefont {Lin},\ and\ \citenamefont {Maiti}}]{Venkatareddy23}%
  \BibitemOpen
  \bibfield  {author} {\bibinfo {author} {\bibfnamefont {N.}~\bibnamefont {Venkatareddy}}, \bibinfo {author} {\bibfnamefont {S.-T.}\ \bibnamefont {Lin}}, \ and\ \bibinfo {author} {\bibfnamefont {P.~K.}\ \bibnamefont {Maiti}},\ }\href@noop {} {\bibfield  {journal} {\bibinfo  {journal} {Phys. Rev. E}\ }\textbf {\bibinfo {volume} {107}},\ \bibinfo {pages} {034607} (\bibinfo {year} {2023})}\BibitemShut {NoStop}%
\bibitem [{\citenamefont {Gonnella}\ \emph {et~al.}(2014{\natexlab{a}})\citenamefont {Gonnella}, \citenamefont {Lamura},\ and\ \citenamefont {Suma}}]{Suma13}%
  \BibitemOpen
  \bibfield  {author} {\bibinfo {author} {\bibfnamefont {G.}~\bibnamefont {Gonnella}}, \bibinfo {author} {\bibfnamefont {A.}~\bibnamefont {Lamura}}, \ and\ \bibinfo {author} {\bibfnamefont {A.}~\bibnamefont {Suma}},\ }\href@noop {} {\bibfield  {journal} {\bibinfo  {journal} {Int. J. Mod. Phys. C}\ }\textbf {\bibinfo {volume} {25}},\ \bibinfo {pages} {1441004} (\bibinfo {year} {2014}{\natexlab{a}})}\BibitemShut {NoStop}%
\bibitem [{\citenamefont {Mie}(1903)}]{Mie1903}%
  \BibitemOpen
  \bibfield  {author} {\bibinfo {author} {\bibfnamefont {G.}~\bibnamefont {Mie}},\ }\href@noop {} {\bibfield  {journal} {\bibinfo  {journal} {Annalen der Physik}\ }\textbf {\bibinfo {volume} {11}},\ \bibinfo {pages} {657} (\bibinfo {year} {1903})}\BibitemShut {NoStop}%
\bibitem [{\citenamefont {Allen}\ and\ \citenamefont {Tildesley}(1989)}]{allen}%
  \BibitemOpen
  \bibfield  {author} {\bibinfo {author} {\bibfnamefont {M.~P.}\ \bibnamefont {Allen}}\ and\ \bibinfo {author} {\bibfnamefont {D.~J.}\ \bibnamefont {Tildesley}},\ }\href@noop {} {\emph {\bibinfo {title} {Computer simulation of liquids}}}\ (\bibinfo  {publisher} {Oxford University Press},\ \bibinfo {year} {1989})\ p.\ \bibinfo {pages} {385}\BibitemShut {NoStop}%
\bibitem [{\citenamefont {Andersen}(1983)}]{rattle}%
  \BibitemOpen
  \bibfield  {author} {\bibinfo {author} {\bibfnamefont {H.~C.}\ \bibnamefont {Andersen}},\ }\href@noop {} {\bibfield  {journal} {\bibinfo  {journal} {J. Comp. Phys.}\ }\textbf {\bibinfo {volume} {52}},\ \bibinfo {pages} {24} (\bibinfo {year} {1983})}\BibitemShut {NoStop}%
\bibitem [{\citenamefont {Plimpton}(1995)}]{plimpton1995fast}%
  \BibitemOpen
  \bibfield  {author} {\bibinfo {author} {\bibfnamefont {S.}~\bibnamefont {Plimpton}},\ }\href@noop {} {\bibfield  {journal} {\bibinfo  {journal} {J. Comp. Phys.}\ }\textbf {\bibinfo {volume} {117}},\ \bibinfo {pages} {1} (\bibinfo {year} {1995})}\BibitemShut {NoStop}%
\bibitem [{\citenamefont {Ester}\ \emph {et~al.}(1996)\citenamefont {Ester}, \citenamefont {Kriegel}, \citenamefont {Sander},\ and\ \citenamefont {Xu}}]{Esler96}%
  \BibitemOpen
  \bibfield  {author} {\bibinfo {author} {\bibfnamefont {M.}~\bibnamefont {Ester}}, \bibinfo {author} {\bibfnamefont {H.-P.}\ \bibnamefont {Kriegel}}, \bibinfo {author} {\bibfnamefont {J.}~\bibnamefont {Sander}}, \ and\ \bibinfo {author} {\bibfnamefont {X.}~\bibnamefont {Xu}},\ }in\ \href@noop {} {\emph {\bibinfo {booktitle} {Proceedings of the Second International Conference on Knowledge Discovery and Data Mining}}}\ (\bibinfo {year} {1996})\BibitemShut {NoStop}%
\bibitem [{\citenamefont {Wojciechowski}\ \emph {et~al.}(1991)\citenamefont {Wojciechowski}, \citenamefont {Frenkel},\ and\ \citenamefont {Bra\'nka}}]{Frenkel91}%
  \BibitemOpen
  \bibfield  {author} {\bibinfo {author} {\bibfnamefont {K.~W.}\ \bibnamefont {Wojciechowski}}, \bibinfo {author} {\bibfnamefont {D.}~\bibnamefont {Frenkel}}, \ and\ \bibinfo {author} {\bibfnamefont {A.~C.}\ \bibnamefont {Bra\'nka}},\ }\href@noop {} {\bibfield  {journal} {\bibinfo  {journal} {Phys. Rev. Lett.}\ }\textbf {\bibinfo {volume} {66}},\ \bibinfo {pages} {3168} (\bibinfo {year} {1991})}\BibitemShut {NoStop}%
\bibitem [{\citenamefont {Wojciechowski}\ \emph {et~al.}(1993)\citenamefont {Wojciechowski}, \citenamefont {Bra\'nka},\ and\ \citenamefont {Frenkel}}]{Frenkel93}%
  \BibitemOpen
  \bibfield  {author} {\bibinfo {author} {\bibfnamefont {K.~W.}\ \bibnamefont {Wojciechowski}}, \bibinfo {author} {\bibfnamefont {A.~C.}\ \bibnamefont {Bra\'nka}}, \ and\ \bibinfo {author} {\bibfnamefont {D.}~\bibnamefont {Frenkel}},\ }\href@noop {} {\bibfield  {journal} {\bibinfo  {journal} {Physica A}\ }\textbf {\bibinfo {volume} {196}},\ \bibinfo {pages} {519} (\bibinfo {year} {1993})}\BibitemShut {NoStop}%
\bibitem [{\citenamefont {Gonnella}\ \emph {et~al.}(2014{\natexlab{b}})\citenamefont {Gonnella}, \citenamefont {Lamura},\ and\ \citenamefont {Suma}}]{gonnella2014phase}%
  \BibitemOpen
  \bibfield  {author} {\bibinfo {author} {\bibfnamefont {G.}~\bibnamefont {Gonnella}}, \bibinfo {author} {\bibfnamefont {A.}~\bibnamefont {Lamura}}, \ and\ \bibinfo {author} {\bibfnamefont {A.}~\bibnamefont {Suma}},\ }\href@noop {} {\bibfield  {journal} {\bibinfo  {journal} {Int. J. Mod. Phys. C}\ }\textbf {\bibinfo {volume} {25}},\ \bibinfo {pages} {1441004} (\bibinfo {year} {2014}{\natexlab{b}})}\BibitemShut {NoStop}%
\bibitem [{\citenamefont {Lei}\ \emph {et~al.}(2019)\citenamefont {Lei}, \citenamefont {{Pica~Ciamarra}},\ and\ \citenamefont {Ni}}]{lei2019nonequilibrium}%
  \BibitemOpen
  \bibfield  {author} {\bibinfo {author} {\bibfnamefont {Q.-L.}\ \bibnamefont {Lei}}, \bibinfo {author} {\bibfnamefont {M.}~\bibnamefont {{Pica~Ciamarra}}}, \ and\ \bibinfo {author} {\bibfnamefont {R.}~\bibnamefont {Ni}},\ }\href@noop {} {\bibfield  {journal} {\bibinfo  {journal} {Science Adv.}\ }\textbf {\bibinfo {volume} {5}},\ \bibinfo {pages} {eaau7423} (\bibinfo {year} {2019})}\BibitemShut {NoStop}%
\end{thebibliography}%

\end{document}